\documentclass[
journal=jctcce,
manuscript=article,
layout=twocolumn
]{achemso}

\makeatletter
\let\l@addto@macro\relax
\makeatother
\usepackage[fontsize=10pt]{scrextend}
\usepackage{bm}
\usepackage[font=small,labelfont=bf]{caption}

\mciteErrorOnUnknownfalse

\usepackage{graphicx}
\usepackage{dcolumn}
\usepackage{bm}
\usepackage{dsfont}
\usepackage{physics}
\usepackage{color}
\usepackage{soul}
\setstcolor{red}
\newcommand{\red}[1]{\textcolor{black}{#1}}
\newcommand{\green}[1]{\textcolor{black}{#1}}
\usepackage{ifthen}
\newboolean{showannotations}   
\setboolean{showannotations}{true} 
\newcommand*{\added}[1]{%
  \ifthenelse{\boolean{showannotations}}{{\color{black}#1}}{#1}%
}

\newcommand*{\addedNEW}[1]{%
  \ifthenelse{\boolean{showannotations}}{{\color{black}#1}}{#1}%
}

 \usepackage[thinc]{esdiff}

\newcommand*{\removed}[1]{%
  \ifthenelse{\boolean{showannotations}}{\st{#1}}{}%
}

\newcommand*{\change}[2]{%
  \ifthenelse{\boolean{showannotations}}{{\color{red}\st{#1}}{\color{blue}#2}}{#2}%
}
\usepackage{geometry}
\usepackage{setspace}
\usepackage{mathrsfs}

\usepackage{amsmath,amssymb,amsfonts,mathtools,amsthm}
\usepackage{braket}
\usepackage{graphicx}
\usepackage{xcolor}
\usepackage{tikz}
\usetikzlibrary{arrows.meta,decorations.pathmorphing,decorations.markings,decorations.pathreplacing,positioning,calc,shapes.geometric}

\author{Amirhosein Amini}
\affiliation{Department of Chemistry, Texas A\&M University, College Station, Texas 77843, USA}
\author{Jaime Cerda}%
\affiliation{Departamento de Física y Astronomía, Facultad de Ciencias Exactas, Universidad Andres Bello, Santiago 837-0136, Chile}
\author{Leopoldo Mejía}%
\affiliation{Departamento de Física y Astronomía, Facultad de Ciencias Exactas, Universidad Andres Bello, Santiago 837-0136, Chile}
\email{leopoldo.mejia@unab.cl}
\author{Arkajit Mandal}%
\affiliation{Department of Chemistry, Texas A\&M University, College Station, Texas 77843, USA}
\email{mandal@tamu.edu}

\title{Many-Body Second Order Green’s Function Theory for  \\ \emph{Ab Initio} Molecular Quantum Electrodynamics}

\begin{document}

\maketitle

\begin{abstract}
{\small
In this work, we develop two many-body quantum electrodynamic methods to calculate the ground-state energies of strongly coupled light-matter molecular systems. Specifically, we extend the second-order many-body Green's function theory (GF2) for electronic systems to incorporate electron-boson couplings. We employ two ansätze to treat the bosonic part of the system, namely the coherent-state (CS) and Lang-Firsov (LF) transformed vacuum state. These are combined with the GF2 method to construct two new approaches, which we refer to as CS-GF2 and LF-GF2.
We benchmark CS- and LF-GF2 by studying various molecular systems inside an optical cavity. We investigate $\mathrm{H}_2$ and $\mathrm{LiH}$ potential energy surfaces, keto-enol tautomerization energy barrier, van der Waals interactions between two $\mathrm{H_2}$ molecules and the torsional potential energy surface of the ethylene molecule, $\mathrm{C_2H_4}$. Both methods provide
\red{similar energy profiles, with relatively small differences between CS-GF2 and LF-GF2. Comparisons with the reference calculations also reveal remaining limitations at stretched bond lengths and near the twisted ethylene configuration.}
\red{We additionally benchmark CS-GF2 against exact diagonalization for a two-site Hubbard-Holstein model and observe that it closely reproduces the coupling-induced energy change in the weakly interacting regime while improving upon CS-MP2 in the strongly interacting regime.}
Overall, the results demonstrate that CS-GF2 and LF-GF2 provide
\red{self-consistent}
descriptions of correlated molecular electron–boson systems, offering a many-body Green’s function alternative for studying ground-state cavity-modified systems.}
\end{abstract}


\begin{tocentry}
\includegraphics[width=\textwidth]{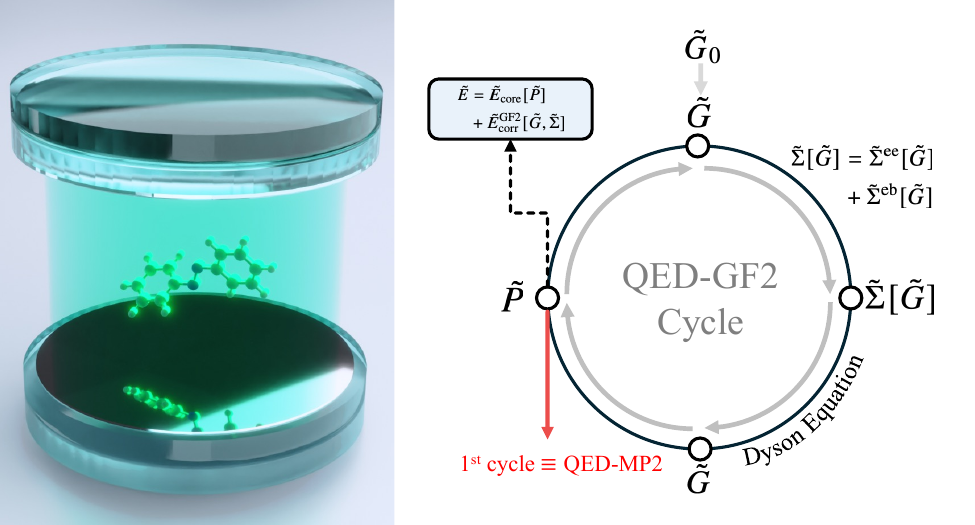}
\end{tocentry}
\section{Introduction}

Experimental work, especially in the past few decades, suggests that strong light-matter interactions  enable a wide range of exotic quantum phenomena~\cite{basov2020polariton, sanvitto2016road, XiangChemRev2024}. These include cavity modification of chemical reactivity~\cite{Nagarajan2021,Xiang2024,Bhuyan2023,Li2022,Ribeiro2018,Khazaei2025,Sharon_proton_transfer, brawley2025vibrational}, cavity enabled fast energy transport~\cite{Zhou2024,Blackham2025,Krupp2025,Balasubrahmaniyam2023,Sandik2024,Xu2023, ying2025microscopic, haines2026mechanistic}, cavity enabled photonic spin-Hall effect~\cite{liang2024polariton, xu2026giant, spencer2021spin}, cavity enhanced electrical conductivity~\cite{Liu2025,Hagenmller2017,Kumar2024}, cavity modification of superconductivity~\cite{Thomas2019,Keren2026,Sentef2018}, and nano and pico-cavity modification of chemistry~\cite{Chang2026, Baumberg2022}. When fully realized, strongly coupled light-matter systems are likely to emerge as a~flexible platform for engineering physicochemical properties of molecular systems beyond the present paradigms of chemistry or physics. 
 
Strong light–matter coupling has been achieved in a variety of platforms, including platforms that couple a single (or a few) molecule to confined radiation~\cite{Chikkaraddy2016, Baumberg2022, santhosh2016vacuum} (i.e. plasmonic nano and pico-cavities), as well as systems where an ensemble of molecules is coupled to an ensemble of radiation modes~\cite{Nagarajan2021, Review_Arkajit, campos2023swinging, borges2025impact, wickramasinghe2026fly} (i.e. Fabry-Pérot cavity). Prior theoretical works~\cite{campos2023swinging, Review_Arkajit, campos2020polaritonic, Lindoy2024, du2023vibropolaritonic, li2020origin} that studied the latter, namely the collective coupling regime, point towards the infeasibility of cavity modification of chemical reactivity. In addition to this, controversy exists with respect to experimental reproducibility or in the interpretation of spectroscopic data for cavity-modified ground-state chemistry in the collective coupling regime~\cite{imperatore2021reproducibility, wiesehan2021negligible, michon2024impact}, and it is presently unclear how collective cavity couplings, which connect cavity radiation to molecular degrees of freedom in a delocalized manner, nevertheless could lead to local modifications of chemical reactivity.

In contrast, a large number of prior theoretical works demonstrate that strong coupling at the single molecular level, while substantially more challenging to achieve experimentally, leads to a modification of both photochemical~\cite{feist2018polaritonic, hoffmann2020effect, fregoni2018manipulating, semenov2019electron, gu2020manipulating, tichauer2021multi, luk2017multiscale, herrera2020molecular, herrera2018theory, mandal2019investigating} and ground state chemical reactivity.~\cite{single_molecule_nature_Flick, Chang2026, vega2025theoretical, sun2022suppression, lindoy2023quantum, li2021cavity, wang2022cavity, Single_Flick, Koch_Nature, Sharon_proton_transfer, garner2025simulation, bauman2025perspective, roden2024perturbative, foley2023ab} At the same time, this single-molecule-strong-coupling regime is  appealing as an ideal testbed for understanding  quantum light–matter interactions at the microscopic level that is theoretically accessible. In contrast to the collective regime, where intractability of large molecular ensembles necessitates the use of  approximate models, this single-molecule limit permits the use of highly accurate many-body
approaches that can resolve microscopic features of light-matter interactions beyond the reach of collective-regime models. This more resolved picture of complex light-matter interactions at the single-molecule limit can provide new insights into the fundamentals at the interface of quantum chemistry and quantum optics.  

The study of strongly coupled light-matter systems falls into two main theoretical frameworks~\cite{weight2025ab, Review_Arkajit, weight2023theory, foley2023ab}: (1) \emph{ab initio} self-consistent quantum electrodynamic (QED) approaches, and (2) parameterized QED approaches. In parameterized QED, the polaritonic Hamiltonian is constructed using adiabatic states obtained by solving the electronic Hamiltonian with electronic structure methods~\cite{Weight2024,Wang2025,pqed_new_huo}. In contrast, \emph{ab initio} self-consistent methods directly extend existing electronic structure theories to incorporate light-matter interactions. Examples include quantum electrodynamic Hartree-Fock (QED-HF)~\cite{PhysRevX.10.041043}, second-order Møller-Plesset perturbation theory (QED-MP2)~\cite{bauer2023perturbation}, Lang-Firsov Møller-Plesset scheme (LF-MP2)~\cite{Cui2024}, Strong Coupling Møller–Plesset perturbation theory (SC-QED-MP2)~\cite{ElMoutaoukal2025}, Density Functional Theory (QED-DFT)~\cite{Ruggenthaler2014,Flick2015}, Diffusion Quantum Monte Carlo approach~\cite{weight2024diffusion}, Coupled Cluster theory (QED-CC)~\cite{PhysRevX.10.041043, Mordovina2020}, Complete Active Space theory (QED-CAS)~\cite{Vu2024, Alessandro2025}, \red{Spin-Flip Configuration Interaction for Strong Static Correlation (QED-SF-CI)~\cite{weight2026spin} }  and (Full) Configuration Interaction (QED-(F)CI)~\cite{Mordovina2020, McTague2022,weight2026spin}. 
Despite advances in \textit{ab initio} QED methods, accurately describing electron-boson-coupled systems remains a challenge. QED-HF offers a computationally efficient but less accurate mean-field approach. QED-DFT’s reliability depends significantly on the choice of functional. Methods such as QED-MP2 and LF-MP2 are limited by their perturbative nature. In contrast, QED-CAS and QED-CCSD can provide high accuracy but at the cost of high computational resources~\cite{bauman2025quantum}.
These limitations highlight the need for theoretical approaches that can simultaneously capture electronic correlation and strong light-matter interactions in a rigorous framework without relying solely on perturbation theory (e.g., MP2-like levels of theory) or inherently expensive methods like CC or CAS, and that can be systematically made more efficient. 

In this work, we approach this problem by developing a many-body Green's function (MBGF) theory for strongly coupled electron-boson systems, such as polaritons. MBGF methods belong to the family of {\it ab initio} self-consistent approaches and provide a natural framework to describe correlation effects and dynamical screening. They have been traditionally used to compute excitation energies in materials, especially within the GW approximation~\cite{hedin1999correlation, shishkin2007self, perfetto2022real}, where electron-electron interactions are treated through a first-order many-body perturbation expansion in the screened Coulomb interaction. More recently, MBGF methods have also been applied to finite systems to compute ground-state energies~\cite{takeshita2019stochastic}, charged excitations~\cite{dou2019stochastic}, and optical excitations~\cite{dou2022time,mejia2023stochastic,mejia2024convergence}. These developments have highlighted the role of second-order many-body perturbation theory in the bare Coulomb interaction, commonly known as GF2, which is particularly useful for small finite systems such as molecules because it includes second-order exchange interactions missing in first-order approximations.

The extension of MBGF methods to strongly coupled light-matter systems is particularly interesting because, first, they provide a systematic route to include correlation effects through many-body perturbation theory and diagrammatic expansions, making the underlying physical approximations transparent~\cite{stefanucci2013nonequilibrium}. Second, they allow electronic and bosonic degrees of freedom to be treated within a unified Green's function framework~\cite{karlsson2021fast, pavlyukh2022time, pavlyukh2022time2}. Third, although MBGF methods are typically computationally demanding, they offer a natural route for algorithmic improvement through their combination with stochastic techniques~\cite{mejia2024convergence, mejia2023stochastic, neuhauser2017stochastic}. Despite these advantages, their utility in the study of strongly coupled light-matter systems remains largely unexplored. Here, we combine coherent-state (CS) and Lang-Firsov (LF) transformations for the bosonic degrees of freedom with GF2 for the electronic many-body problem, leading to two {\it ab initio} QED approaches: CS-GF2 and LF-GF2. We benchmark these methods across several molecular systems and
\red{assess the changes introduced by GF2 self-consistency, including the remaining discrepancies with exact and active-space reference calculations.}

\section{Theory}

\subsection{Light-Matter Hamiltonian}\label{LM Hamiltonian}
We consider a molecular system coupled to a quantized electromagnetic field inside an optical cavity, and work in a non-orthogonal atomic-orbital basis $\{ \varphi_i(\mathbf{r}) \}$ with overlap matrix $S_{ij} = \langle \varphi_i | \varphi_j \rangle$. The coupled electron-boson system can be described via a general light-matter Hamiltonian of the form~\cite{weight2023theory, bauman2025perspective, Review_Arkajit, Ruggenthaler2023}
\begin{equation}
    \begin{aligned}
\hat{H}
=
\sum_{ij} h_{ij} a_i^\dagger a_j
&+
\frac{1}{2}\sum_{ijkl} v_{ijkl} a_i^\dagger a_k^\dagger \red{a_l a_j}\\
&+
\omega_c b^\dagger b
+
\sum_{ij} g_{ij}\, a_i^\dagger a_j (b+b^\dagger),
\end{aligned}
\label{LM Hamil}
\end{equation}
where $a_i^\dagger$ ($a_i$) is the electronic creation (annihilation) operator in orbital $i$, $b^\dagger$ ($b$) is the bosonic creation (annihilation) operator corresponding to the cavity mode with frequency $\omega_c$, and $g_{ij}$ are matrix elements of the electron-boson coupling.
The one- and two-electron integrals $h_{ij}$ and $v_{ijkl}$ are defined as
\begin{equation}
    \label{hij}
    h_{ij} = \int \mathrm{d}\mathbf{r} \: \varphi_j(\mathbf{r}) \left( -\frac{1}{2} \nabla^2 + v_{\text{nuc}} \right) \varphi_i(\mathbf{r}),
\end{equation}
and
\begin{equation}
    \label{vijkl}
    v_{ijkl} = \iint \varphi_i(\mathbf{r}) \varphi_j(\mathbf{r}) \dfrac{1}{|\mathbf{r} - \mathbf{r}'|} \varphi_k(\mathbf{r}') \varphi_l(\mathbf{r}') \, \mathrm{d}\mathbf{r} \, \mathrm{d}\mathbf{r}'.
\end{equation}
\paragraph{Pauli-Fierz Hamiltonian}
In this work, we treat the light-matter interaction within the long-wavelength approximation. Extensions beyond this approximation are possible and are left for future work. This interaction is described by the \emph{Pauli-Fierz} Hamiltonian~\cite{Review_Arkajit, rokaj2018light} within the dipole gauge. The electron-photon coupling matrix elements $g_{ij}$ are given by
\begin{equation}
    \label{g-ij}
    g_{ij} = - \sqrt{\dfrac{\omega_c}{2}}\eta_c   \boldsymbol{\varepsilon} \cdot \boldsymbol{\mu}_{ij} = - \sqrt{\dfrac{\omega_c}{2}}\eta_c  {\mu}_{ij},
\end{equation}
where $\eta_c$ is the light-matter coupling strength, $\boldsymbol{\varepsilon}$ is the photonic polarization unit vector, and $\boldsymbol{\varepsilon}  \cdot  \boldsymbol{\mu}_{ij} = {\mu}_{ij}$ is the molecular dipole integral given by
\begin{equation}
    \label{MOL DIPOLE}
    \mathbf{\mu}_{ij} = 	\int d\mathbf{r}\,
	\varphi_i(\mathbf{r})\, \mathbf{r}\, \varphi_j(\mathbf{r})
	-
	\sum_A Z_A \mathbf{R}_A\, S_{ij},
\end{equation}
where $Z_A$ and $\mathbf{R}_A$ are the nuclear charge and the position of atom $A$. 
\paragraph{Dipole Self-energy}
Within the dipole gauge approximation, the Pauli-Fierz Hamiltonian also includes the \emph{dipole self-energy} term of the cavity mode~\cite{Cui2024}
\begin{equation}
    \label{DSE}
    \text{DSE} 
	=
	\frac{\eta_c^2}{2}
	\sum_{ijkl}
	(\boldsymbol{\varepsilon}\cdot \mu_{ij})
	(\boldsymbol{\varepsilon}\cdot \mu_{kl})
	a_i^\dagger a_j a_k^\dagger a_l.
\end{equation}
In the second-quantization formalism, the addition of the DSE modifies the one- and two-electron integrals~\cite{Cui2024}, and one can write the new \emph{effective} integrals as
\begin{equation}
    \label{h_eff}
    	h_{ij}^{\mathrm{eff}}
	=
	h_{ij}
	+
	\frac{\eta_c^2}{2}
	\sum_{k}
	(\boldsymbol{\varepsilon}\cdot \boldsymbol{\mu})_{ik}
	(\boldsymbol{\varepsilon}\cdot \boldsymbol{\mu})_{kj},
\end{equation}
and
\begin{equation}
    v_{ijkl}^{\text{eff}} = v_{ijkl} + \eta_c^2 (\boldsymbol{\varepsilon} \cdot \boldsymbol{\mu})_{ij} (\boldsymbol{\varepsilon} \cdot \boldsymbol{\mu})_{kl}.
\end{equation}
Replacing the electron integrals in Eq.~\eqref{LM Hamil} with their effective form and defining $X \equiv b + b^\dagger$ for convenience, one gets
\begin{equation}
    \begin{aligned}
    \hat{H}
	=
	\sum_{ij} h_{ij}^{\mathrm{eff}} a_i^\dagger a_j
	&+
	\frac{1}{2}\sum_{ijkl} v_{ijkl}^{\mathrm{eff}} a_i^\dagger a_k^\dagger \red{a_l a_j}\\
	&+
	\omega_c\, b^\dagger b
	+
	\sum_{ij} g_{ij}\, a_i^\dagger a_j X.
    \end{aligned}
	\label{eq:H_QED_final}
\end{equation}

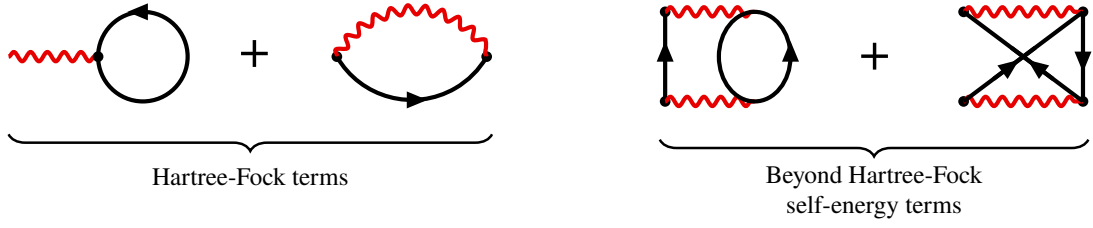
\begin{figure*}
    \centering
    \resizebox{0.8\textwidth}{!}{%
\begin{tikzpicture}[
line width=2.0pt,
fermion/.style={postaction={decorate},decoration={markings,mark=at position #1 with {\node[transform shape, fill=black, inner sep=1.4pt, draw, isosceles triangle]{};}}},
fermion/.default=0.5,
photon/.style={red!90!black,decorate,decoration={snake, amplitude=2.2pt, segment length=8pt}},
vertex/.style={circle, fill=black, inner sep=1.8pt},
arrow/.style={postaction={decorate},decoration={markings,mark=at position #1 with {\node[transform shape, fill=black, inner sep=1.4pt, draw, isosceles triangle]{};}}},
plus/.style={font=\Huge, text=black},
groupbrace/.style={decorate,decoration={brace, mirror, amplitude=8pt},line width=1.2pt},
grouplabel/.style={font=\large,align=center,yshift=-20pt}
]
\newcommand{\fermioncircle}[3]{\draw[arrow=#3] #1 circle[radius=#2];}
\newcommand{\fermionellipse}[4]{\draw[arrow=#4] (#1) ellipse[x radius=#2, y radius=#3];}

\begin{scope}[local bounding box=d1]
\draw[photon] (0,0) -- (1.5,0);
\node[vertex] at (1.5,0) {};
\fermioncircle{(2.25,0)}{0.75}{0.25}
\end{scope}
\node[right=0.7cm of d1, plus] {$+$};

\begin{scope}[xshift=5.5cm, local bounding box=d2]
\coordinate (A) at (0,0);
\coordinate (B) at (2.51,0);
\node[vertex] at (A) {};
\node[vertex] at (B) {};
\draw[photon] (A) arc[start angle=150, end angle=30, radius=1.45];
\draw[arrow=0.5] (A) arc[start angle=-150, end angle=-30, radius=1.45];
\end{scope}

\begin{scope}[xshift=11cm, yshift=-0.75cm, local bounding box=d3]
\coordinate (TL) at (0,1.5);
\coordinate (BL) at (0,0);
\coordinate (TR) at (1.5,1.5);
\coordinate (BR) at (1.5,0);
\node[vertex] at (TL) {};
\node[vertex] at (BL) {};
\draw[fermion=0.5] (BL) -- (TL);
\draw[photon] (TL) -- (TR);
\draw[photon] (BL) -- (BR);
\fermionellipse{$(TR)!0.5!(BR)$}{0.6}{0.75}{0.0}
\end{scope}
\node[right=0.9cm of d3, plus] {$+$};

\begin{scope}[xshift=16cm, yshift=-0.75cm, local bounding box=d4]
\coordinate (TL4) at (0,1.5);
\coordinate (BL4) at (0,0);
\coordinate (TR4) at (2,1.5);
\coordinate (BR4) at (2,0);
\node[vertex] at (TL4) {};
\node[vertex] at (BL4) {};
\node[vertex] at (TR4) {};
\node[vertex] at (BR4) {};
\draw[photon] (TL4) -- (TR4);
\draw[photon] (BL4) -- (BR4);
\draw[fermion=0.5] (TR4) -- (BR4);
\draw[fermion=0.6] (BL4) -- ($(BL4)!0.6!(TR4)$);
\draw (0.5,0.375) -- (TR4);
\draw[fermion=0.6] (BR4) -- ($(BR4)!0.6!(TL4)$);
\draw (1.5,0.375) -- (TL4);
\end{scope}

\draw[groupbrace]
([yshift=-0.55cm] d1.south west) --
([yshift=-0.55cm] d2.south east |- d1.south west)
node[midway, grouplabel] {Hartree-Fock terms};
\draw[groupbrace]
($(d3.south west)+(0,-0.55)$) --
($(d4.south east)+(0,-0.55)$)
node[midway, grouplabel, yshift=-5pt] {Beyond Hartree-Fock\\self-energy terms};
\end{tikzpicture}%
}
\caption{Self-energy diagrams included in the GF2 approximation, where wavy lines denote two-electron interaction lines and solid directed lines denote electronic Green's functions. The first two diagrams correspond to the Hartree-Fock contributions, while the last two diagrams (Eq.~\ref{eq:Sigma_GF2}) represent the beyond-Hartree-Fock second-order correlation terms.}
    \label{Sigma_Elec}
\end{figure*}


\subsection{GF2}\label{GF2_Elec}
In this section, we review the electronic GF2 formalism that provides the basis for the polaritonic GF2 derivation in Sec.~\ref{polGF2}. The method is formulated in the Matsubara Green's function framework, which enables the evaluation of finite-temperature expectation values within the grand canonical ensemble. In the calculations presented here, we consider the low-temperature limit, so that Matsubara GF2 effectively functions as a ground-state electronic structure method.

\paragraph{Matsubara Green's function}

In many-body Green's function methods, the central object is the one-particle Green's function of the system. For electrons in thermal equilibrium, the Matsubara Green's function is defined as
\begin{equation}
	G_{ij}(\tau) \equiv
	-\langle \mathscr{T}\, a_i(\tau)\, a_j^\dagger(0)\rangle,
	\qquad \tau\in(-\beta,\beta),
	\label{eq:G_def}
\end{equation}
where the imaginary-time Heisenberg operators are
$a_i(\tau)=e^{\tau \hat{K}}a_i e^{-\tau \hat{K}}$ and
$a_j^\dagger(0)=a_j^\dagger$, with
$\hat{K}\equiv \hat{H}-\mu\hat{N}$. Here, $\mathscr{T}$ denotes the fermionic imaginary-time ordering operator. Because $\hat{K}$ is time independent, the equilibrium Green's function depends only on the time difference, allowing us to set the second time argument to zero. It is often convenient to work with the Fourier representation of the Green's function
\begin{equation}
    \mathbf{G}(i\omega_n) = \int_0^\beta d\tau\, e^{i\omega_n\tau}\,
	\mathbf{G}(\tau),
\end{equation}
such that
\begin{equation}
    \mathbf{G}(\tau) = \frac{1}{\beta}\sum_{n=-\infty}^{\infty}
	e^{-i\omega_n \tau}\,\mathbf{G}(i\omega_n),
    	\label{eq:Fourier_G}
\end{equation}
where $\omega_n = (2n+1)\pi/\beta$, $n\in\mathbb{Z}$ are the fermionic Matsubara frequencies. Taking the limit
$\tau\to 0^-$ in Eq.~\eqref{eq:G_def} gives
$G_{ij}(0^-) = \langle a_j^\dagger a_i\rangle$, so that for a
closed-shell system, the one-particle reduced density matrix is obtained
directly from the Matsubara sum~\cite{neuhauser2017stochastic},
\begin{equation}
\begin{aligned}
P_{ij} &= \red{2 G_{ij}(0^-)} = -2 G_{ij}(\beta^-)\\
	& = \frac{4}{\beta}\,\mathrm{Re}
	\sum_{n=0}^{\infty} e^{-i\omega_n 0^-}\, G_{ij}(i\omega_n),
\end{aligned}
	\label{eq:P_from_G}
\end{equation}
where the last equality assumes a real-valued representation of $\mathbf{G}(\tau)$ on $(0,\beta)$, as is the case for the real molecular Hamiltonians considered here.

\paragraph{Dyson equation and self-energy}
By defining a \red{mean-field reference} Green's function $\mathbf{G}_0$ and using perturbation theory, we can build approximations for $\mathbf{G}(\tau)$.
We take as a reference the Hartree-Fock Hamiltonian
$\hat{H}_0 = \sum_{ij} F_{ij}\,a_i^\dagger a_j$ in which the Fock matrix,
\begin{equation}
	F_{ij} = h_{ij} + \frac{1}{2}\sum_{kl}P_{kl}
\left(2 v_{ijkl} - v_{ilkj}\right),
	\label{eq:FockMat}
\end{equation}
is built self-consistently from the correlated density matrix.

The corresponding \red{mean-field reference} Matsubara Green's function can be
obtained by transforming to an orthogonal basis via a matrix $\mathbf{X}$
satisfying $\mathbf{X}\mathbf{X}^T = \mathbf{S}^{-1}$, diagonalizing
$\overline{\mathbf{F}} = \mathbf{X}^T\mathbf{F}\mathbf{X}$, and
rotating back to the AO representation~\cite{dahlen2005self,phillips2014communication,neuhauser2017stochastic}. The result can be shown as the compact equation
\begin{equation}
	\mathbf{G}_0(i\omega_n) =
	\bigg[(\mu+i\omega_n)\mathbf{S} - \mathbf{F}\bigg]^{-1}.
	\label{eq:G0}
\end{equation}
The reference Green's function $\mathbf{G}_0$ contains the one-body propagation generated by the correlated Fock matrix and therefore accounts for interactions only at the mean-field level. Correlation effects beyond the mean-field can be accounted for using the \emph{self-energy} $\mathbf{\Sigma}$, defined as
\begin{equation}
	\mathbf{\Sigma}(i\omega_n) \equiv
	\mathbf{G}_0^{-1}(i\omega_n) - \mathbf{G}^{-1}(i\omega_n),
\end{equation}
which immediately yields the Dyson equation~\cite{neuhauser2017stochastic}, written in two equivalent forms,
\begin{align}
	\mathbf{G}(i\omega_n) &=
	\mathbf{G}_0(i\omega_n) + \mathbf{G}_0(i\omega_n)\,
	\mathbf{\Sigma}(i\omega_n)\,\mathbf{G}(i\omega_n), \\
	\mathbf{G}(i\omega_n) &=
	\left[(\mu+i\omega_n)\mathbf{S} - \mathbf{F}
	- \mathbf{\Sigma}(i\omega_n)\right]^{-1}.
	\label{eq:Dyson}
\end{align}

\red{
The partitioning of the self-energy depends on the choice of reference propagator. Relative to the bare one-electron Hamiltonian, the electronic self-energy through second order can be written schematically as
\begin{equation}
    \mathbf{\Sigma}_{\mathrm{total}}
    =
    \mathbf{\Sigma}_{\mathrm{H}}
    +
    \mathbf{\Sigma}_{\mathrm{F}}
    +
    \mathbf{\Sigma}^{(2)},
\end{equation}
where $\mathbf{\Sigma}_{\mathrm{H}}$ and $\mathbf{\Sigma}_{\mathrm{F}}$ denote the static first-order Hartree and Fock contributions, respectively. In the present formulation, these first-order contributions are already incorporated into the self-consistent Fock matrix in Eq.~\eqref{eq:FockMat}. Consequently, because the Dyson equation in Eq.~\eqref{eq:Dyson} is formulated with respect to the Fock-reference propagator $\mathbf{G}_0$, the self-energy $\mathbf{\Sigma}$ appearing explicitly in the Dyson equation contains only the beyond-Hartree-Fock correlation contribution.
}
\red{
Within GF2, this correlation self-energy is approximated by the second-order direct and exchange skeleton diagrams in the bare Coulomb interaction, evaluated self-consistently with the interacting Green's function~\cite{baym1961conservation,baym1962self,phillips2014communication}:
}
\begin{equation}
\begin{aligned}
    	\Sigma_{ij}(\tau) = &\sum_{klmnpq}-G_{kl}(\tau)G_{mn}(\tau)
	G_{pq}(\beta-\tau) \\
    &\times v_{impk}\left(2 v_{jnlq} - v_{jlnq}\right).
\end{aligned}
	\label{eq:Sigma_GF2}
\end{equation}
The diagrammatic form of the electronic many-body interactions considered in this approximation is shown in Fig. \ref{Sigma_Elec}.
\red{
The first two diagrams represent the static Hartree and Fock contributions already incorporated into the self-consistent Fock matrix, whereas only the final two diagrams correspond to the second-order correlation self-energy in Eq.~\eqref{eq:Sigma_GF2}.
}
%


 \begin{figure*}
\centering
\includegraphics[width=0.84\linewidth]{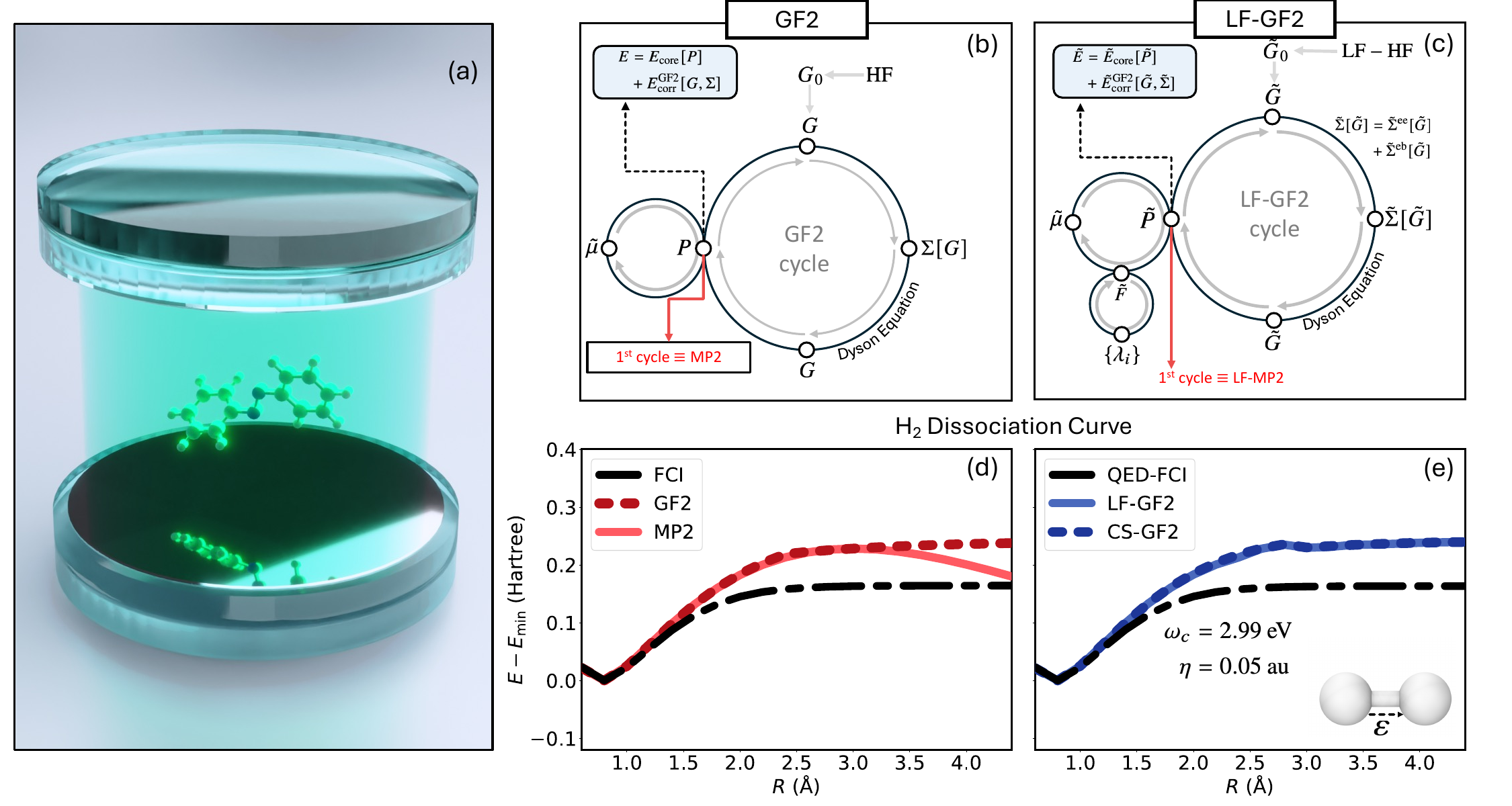}
\caption{\footnotesize
Schematic representation of the self-consistent procedures and $\mathrm{H_2}$ dissociation curves inside and outside an optical cavity.
(a) Illustration of a molecule inside an optical cavity.
(b) Standard electronic GF2 workflow: starting from a converged Hartree-Fock calculation, the Fock matrix $F_{ij}$ and density matrix $P_{ij}$ are used to construct the Green's function $G_{ij}$ (Eq.~\ref{eq:G0}), from which the self-energy $\Sigma_{ij}[G_{ij}]$ is evaluated (Eq.~\ref{eq:Sigma_GF2}), enabling iterative solution of the Dyson equation (Eq.~\ref{eq:Dyson}); the first iteration is equivalent to MP2.
(c) LF-GF2 workflow: a Lang-Firsov (LF) transformation introduces additional variational parameters $\{\lambda_i\}$ and QED-dressed quantities $\tilde{P}_{ij}, \tilde{F}_{ij}, \tilde{G}_{ij}$; electronic ($\Sigma^{ee}$) and electron-boson ($\Sigma^{eb}$) contributions are included self-consistently, with the first iteration corresponding to LF-MP2.
(d) Dissociation curve of $\mathrm{H_2}$ outside the cavity.
\red{GF2 removes the downward turn of the MP2 curve at stretched bond lengths, but remains above the FCI relative-energy curve.}
(e) Dissociation curve inside the cavity for $\omega_c = 2.99~\mathrm{eV}$ and $\eta = 0.05\:\mathrm{au}$,
\red{showing nearly coincident CS-GF2 and LF-GF2 profiles and their deviation from QED-FCI at stretched bond lengths.}
The energy calculations in panels (d) and (e) used the cc-pVDZ basis set.
}
\label{Schematic Procedure and H2 Figure}
\end{figure*}


The total electronic energy is then computed using ~\cite{Galitskii1958,baym1961conservation,baym1962self,dahlen2005self,neuhauser2017stochastic,takeshita2019stochastic,rusakov2016self,phillips2014communication}
\begin{equation}
	E_{\mathrm{GF2}} =
	\underbrace{
	\frac{1}{2}\,\mathrm{Tr}\!\left[
	(\mathbf{h}+\mathbf{F})\mathbf{P}\right]
	}_{E^{\mathrm{Hartree\text{-}Fock}}}
	-\int_0^\beta d\tau\,
    \mathrm{Tr}\left[
    \mathbf{\Sigma}^{ee}(\tau)\mathbf{G}(\beta-\tau)
    \right],
	\label{eq:E_GF2}
\end{equation}
or, equivalently, the frequency-domain form
\begin{equation}
\begin{aligned}
    E_{\mathrm{GF2}} = 
	&E^{\mathrm{Hartree\text{-}Fock}}  \\
	&+ \frac{2}{\beta}\,\mathrm{Re}
	\sum_{n=0}^{\infty} e^{-i\omega_n 0^-}\,
	\mathrm{Tr}\!\left[\mathbf{\Sigma}(i\omega_n)\,
	\mathbf{G}(i\omega_n)\right],
\end{aligned}
	\label{eq:E_GF2_freq}
\end{equation}
The first term in Eqs.~\eqref{eq:E_GF2} and~\eqref{eq:E_GF2_freq} corresponds to the mean-field energy evaluated on the correlated density matrix $\mathbf{P}$, while the second term is the GF2 correlation energy, which follows from the \emph{Galitskii-Migdal} identity~\cite{Galitskii1958}.

\paragraph{Self-Consistent Procedure}
Equations~\eqref{eq:P_from_G}, \eqref{eq:FockMat}, \eqref{eq:Dyson}, and \eqref{eq:Sigma_GF2}, together with the energy expression Eq.~\eqref{eq:E_GF2}, define a closed self-consistent scheme on a Matsubara grid. Starting from a converged Hartree--Fock solution with $\mathbf{\Sigma}(i\omega_n) = \mathbf{0}$, one iterates the following steps until convergence of the total energy:
\begin{enumerate}
    \item Given $\mathbf{F}$ and $\mathbf{\Sigma}(i\omega_n)$, build $\mathbf{G}(i\omega_n)$ from Dyson's equation~\eqref{eq:Dyson} and transform to imaginary time to obtain $\mathbf{G}(\tau)$.
    \item Update the density matrix from Eq.~\eqref{eq:P_from_G}, adjusting the chemical potential $\mu$ to enforce the electron-number constraint $\mathrm{Tr}[\mathbf{P}\mathbf{S}] = N_e$.
    \item Rebuild the Fock matrix from Eq.~\eqref{eq:FockMat}.
	\item Compute the self-energy in imaginary time from Eq.~\eqref{eq:Sigma_GF2} and transform back to frequency domain to obtain $\mathbf{\Sigma}(i\omega_n)$.
\end{enumerate}
The computational cost is dominated by the self-energy construction in Eq.~\eqref{eq:Sigma_GF2}, which scales as $\mathcal{O}(N_\tau N^5)$ for $N$ AO basis functions and $N_\tau$ imaginary-time grid points.
\green{However, the cost of the electronic GF2 self-energy can be reduced through stochastic evaluation of the self-energy contractions, as demonstrated in stochastic GF2 and stochastic resolution-of-identity GF2 formulations~\cite{neuhauser2017stochastic,takeshita2019stochastic, mejia2023stochastic, mejia2024convergence}. Extending these stochastic approaches to the electronic and electron-boson self-energies of QED-GF2 provides a promising route for reducing the computational cost of the present method.}


\red{When the GF2 self-energy is evaluated once using the Hartree-Fock Green's function, the MP2 correlation energy can be obtained in the zero-temperature limit as~\cite{neuhauser2017stochastic} \begin{equation}
E_{\mathrm{MP2}}^{\mathrm{corr}}
=
-\frac{1}{2}\int_0^\beta d\tau\,
\mathrm{Tr}\!\left[
\mathbf{\Sigma}^{ee}[\mathbf{G}_0](\tau)
\mathbf{G}_0(\beta-\tau)
\right],
\quad \beta\rightarrow\infty.
\label{eq:MP2_reference_contraction}
\end{equation}}

Therefore, GF2 can be viewed as a self-consistent extension of MP2, where each iterative update of $\mathbf{G}$ effectively incorporates higher-order correlation effects beyond second-order perturbation theory. The GF2 self-consistent procedure is schematically illustrated in Fig.~\ref{Schematic Procedure and H2 Figure}(b).


\subsection{QED-GF2}\label{polGF2}
We now extend the electronic GF2 formalism of Sec.~\ref{GF2_Elec} to coupled electron-boson systems described by the light--matter Hamiltonian in Eq.~\eqref{eq:H_QED_final}. The resulting QED-GF2 method uses a QED-dressed mean-field reference and incorporates beyond-mean-field effects through two self-energy contributions: the electronic GF2 self-energy, $\Sigma^{ee}$, and the electron-boson Fan self-energy, $\Sigma^{eb}$, as detailed below. The electronic propagator is therefore dressed by both electron-electron and electron-boson correlations, while the bosonic propagator is dressed by the electronic polarization $\Pi$.

Following the electronic GF2 construction, we start from a QED-dressed mean-field reference. Depending on the transformation used for the bosonic degrees of freedom, this reference corresponds either to the coherent-state Hartree--Fock (CS-HF) or Lang--Firsov Hartree--Fock (LF-HF) solution. We denote the corresponding QED-dressed Fock matrix by $\tilde{\mathbf{F}}$ and the associated reference energy by $\tilde{E}$. The reference Hamiltonian contains the one-body electronic terms generated by the chosen QED mean-field theory, while the residual electron-electron and electron-boson interactions are treated perturbatively through self-energy corrections.
\red{The following expressions are first written for the CS partition. The coherent displacement is included in the static reference, and $X$ denotes the remaining fluctuation coordinate, such that $\langle X\rangle=0$. Therefore writing $\hat Q=\sum_{ij\sigma}g_{ij}a_{i\sigma}^\dagger a_{j\sigma}$ and $A=\mathrm{Tr}[\mathbf g\mathbf P]$, the residual linear interaction is $(\hat Q-A)X$. The corresponding LF interaction and residual vertex are specified below.}

\paragraph{Bosonic Green's Function}
We define the bosonic displacement Green's function as
\begin{equation}
    D(\tau) \equiv - \langle \mathscr{T}\, X(\tau)\, X(0) \rangle,
    \label{Boson Green's Function Def}
\end{equation}
where we have taken advantage of the time-translation invariance to set the second imaginary time argument equal to zero. Because $D(\tau+\beta)=D(\tau)$ following KMS boundary conditions\cite{haag1967equilibrium, kubo1957statistical,martin1959theory}, we can write the Fourier expansion of the bosonic Green's function as
\begin{equation}
D(\tau)=\frac{1}{\beta}\sum_\ell e^{-i\nu_\ell\tau}D(i\nu_\ell),
\label{eq:fourier_boson_qed}
\end{equation}
with
\begin{equation}
    D(i\nu_\ell)=\int_0^\beta d\tau\, e^{i\nu_\ell\tau}D(\tau),
    \label{Dinuell}
\end{equation}
and bosonic Matsubara frequencies $\nu_\ell=2\pi \ell/\beta$, $\ell \in \mathbb{Z}$.

The corresponding noninteracting Green's function is given by~\cite{giustino2017electron}
\begin{equation}
    \label{Non-interacting-D0}
    D_0(i \nu_\ell) = -\dfrac{2 \omega_c}{\nu_\ell^2 + \omega_c^2},
\end{equation}
while beyond-mean-field corrections to the bosonic propagator enter through the \emph{bosonic polarization} $\Pi$~\cite{Stefanucci2013, mahan2000many}:
\begin{equation}
    D^{-1}(i\nu_\ell) = D_0^{-1}(i\nu_\ell) - \Pi(i \nu_\ell),
    \label{Bosonic Dyson}
\end{equation}
which we refer to as the bosonic Dyson equation~\cite{rainer1986principles}.

\begin{figure}[h]
    \centering
\begin{tikzpicture}[
line width=2.0pt,
fermion/.style={
    postaction={decorate},
    decoration={
        markings,
        mark=at position #1 with {
            \node[transform shape, fill=black, inner sep=1.4pt, draw,
            isosceles triangle]{};
        }
    }
},
fermion/.default=0.5,
photon/.style={
    blue!90!black,
    decorate,
    decoration={snake, amplitude=2.2pt, segment length=8pt}
},
externalboson/.style={
    blue!90!black,
    line width=1.1pt,
    decorate,
    decoration={coil, amplitude=1.0pt, segment length=5pt}
},
externalfermion/.style={
    black,
    line width=1.0pt
},
vertex/.style={circle, fill=black, inner sep=1.8pt},
arrow/.style={
    postaction={decorate},
    decoration={
        markings,
        mark=at position #1 with {
            \node[transform shape, fill=black, inner sep=1.4pt, draw,
            isosceles triangle]{};
        }
    }
}
]

\begin{scope}[local bounding box=fan]
\draw[externalfermion] (0.8,0) -- (1.5,0);
\draw[externalfermion] (3.0,0) -- (3.7,0);

\node[vertex] (v1) at (1.5,0) {};
\node[vertex] (v2) at (3.0,0) {};

\draw[fermion=0.5] (v1) -- (v2);
\draw[photon] (v1) to[out=65,in=115] (v2);

\node at (2.25,-0.28) {$\tilde{G}$};
\node at (2.25,0.95) {$D$};
\node at (1.35,-0.35) {$g$};
\node at (3.15,-0.35) {$g$};
\end{scope}

\node at (2.25,-0.9) {$\Sigma^{\mathrm{eb}}$ (Fan) $\sim \red{-}g\tilde{G}gD$};
\begin{scope}[yshift=-2.25cm, local bounding box=bubble]
\node[vertex] (w1) at (1.5,0) {};
\node[vertex] (w2) at (3.0,0) {};

\draw[externalboson] (0.9,0) -- (w1);
\draw[externalboson] (w2) -- (3.6,0);

\draw[arrow=0.5] (w1) to[out=70,in=110] (w2);
\draw[arrow=0.5] (w2) to[out=-110,in=-70] (w1);

\node at (2.25,0.85) {$\tilde{G}$};
\node at (2.25,-0.85) {$\tilde{G}$};

\node at (1.45,0.28) {$g$};
\node at (3.05,0.28) {$g$};

\end{scope}

\node at (2.25,-3.58) {$\Pi \sim -2g\tilde{G}g\tilde{G}$};

\end{tikzpicture}
\caption{
Diagrammatic representation of the electron-boson self-energy $\Sigma^{\mathrm{eb}}$ (Eq.~\eqref{eq:Sigma_eb_freq_qed}) and the bosonic polarization $\Pi$ (Eq.~\eqref{eq:Pi_time_qed}) entering the QED-GF2 approximation. Upper: Electron-boson self-energy (Fan contribution), 
$\Sigma^{\mathrm{eb}}\sim \red{-}g\tilde{G}gD$. The blue wavy line denotes the bosonic Green's function $D$ (Eq.~\eqref{Dinuell}), the thick solid black line denotes the QED-dressed electronic Green's function $\tilde{G}$, and the black dots denote electron-boson interaction vertices associated with the coupling $g$. Lower: Bosonic polarization, $\Pi \sim -2g\tilde{G}g\tilde{G}$, represented by an electron-hole bubble. The blue coil lines denote external bosonic legs, indicating that $\Pi$ enters the bosonic Dyson equation.
}
    \label{fig:Sigma_eb}
\end{figure}
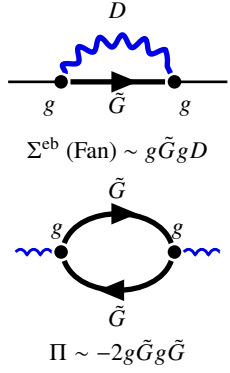

\paragraph{Electron-boson Self-energy}
In addition to the purely electronic self-energy discussed in Sec.~\ref{GF2_Elec}, the interaction between electrons and bosons introduces an additional self-energy. At second order, the dynamical contribution to the fermionic self-energy from bosons is given by the electron-boson \emph{Fan} or \emph{Fan-Migdal}-type self-energy (see a detailed derivation in the Supporting Information)~\cite{fan1951temperature,becker1949optical,giustino2017electron},
\begin{equation}
\Sigma^{eb}_{ij}(\tau)
=
\red{-\sum_{kl}}
g_{ik}\,G_{kl}(\tau)\,g_{lj}\,D(\tau),
\qquad 0<\tau<\beta.
\label{eq:Sigma_eb_time_qed}
\end{equation}
This describes the exchange of a virtual boson between electronic states. In Matsubara frequency space, this expression becomes~\cite{Galitskii1958,engelsberg1963coupled,fetter2003quantum, mahan2000many, giustino2017electron}
\begin{equation}
\Sigma^{eb}_{ij}(i\omega_n)
=
\red{-\frac{1}{\beta}}\sum_{\ell}\sum_{kl}
g_{ik}\,G_{kl}(i\omega_n-i\nu_\ell)\,g_{lj}\,D(i\nu_\ell),
\label{eq:Sigma_eb_freq_qed}
\end{equation}
 which is represented diagrammatically in Fig.~\ref{fig:Sigma_eb}~\cite{mahan2000many, Stefanucci2013, fetter2003quantum,giustino2017electron}.

To evaluate the Fan self-energy, one needs to be able to calculate $D(\tau)$ or, equivalently, $D(i \nu_\ell)$ through Eq.~\eqref{Bosonic Dyson}. The corresponding bosonic polarization, which enters the bosonic Dyson equation, is obtained as~\cite{hedin1970effects, mahan2000many, giustino2017electron}
 \begin{equation}
  \Pi(\tau)
  =
  -2\sum_{ij}\sum_{kl}
  g_{ij}\,G_{jk}(\tau)\,g_{kl}\,G_{li}(\beta-\tau),
  \label{eq:Pi_time_qed}
 \end{equation}
 or equivalently
\begin{equation}
\Pi(i\nu_\ell)
=
\red{\frac{2}{\beta}}
\sum_n\sum_{ij}\sum_{kl}
g_{ij}\,G_{jk}(i\omega_n)\,g_{kl}\,
G_{li}(i\omega_n-i\nu_\ell).
\label{eq:Pi_freq_qed}
\end{equation}
    Together, Eqs. ~\eqref{eq:Sigma_eb_time_qed}-\eqref{eq:Pi_freq_qed} describe the dynamical corrections originating from the electron-boson interactions, completing the set of relevant self-energies for the coupled system. The total electronic self-energy is therefore given by
\begin{equation}
\mathbf{\tilde{\Sigma}}(i\omega_n)
=
\mathbf{\Sigma}^{ee}(i\omega_n)
+
\mathbf{\Sigma}^{eb}(i\omega_n),
\label{total_sigma}
\end{equation}
which defines the QED-GF2 approximation. Here, $\Sigma^{ee}$ has the same second-order GF2 form as Eq.~\eqref{eq:Sigma_GF2}, but is evaluated using the QED-dressed electronic Green's function $\tilde{G}$.

The electronic Dyson equations of QED-GF2 is then written as
\begin{equation}
\begin{aligned}
&\mathbf{\tilde{G}}(i\omega_n) =
\bigg[
(\mu+i\omega_n)\mathbf{S}
-\mathbf{\tilde{F}}
-\mathbf{\Sigma}^{ee}(i\omega_n) 
-\mathbf{\Sigma}^{eb}(i\omega_n) \bigg]^{-1},
\end{aligned}
 \label{eq:Dyson_total}
\end{equation}
which must be solved self-consistently together with Eq.\eqref{Bosonic Dyson}. 

\red{After convergence, the total energy is evaluated by including the static reference contribution, which is evaluated with the correlated density matrix, the electronic correlation contribution, the residual electron-boson interaction energy, and the bosonic fluctuation energy~\cite{saekkinen2015manybody}. For the CS partition, this contribution is \begin{equation} \begin{aligned} E_{\mathrm{stat}}^{\mathrm{CS}} ={}& \mathrm{Tr}[\mathbf h^{\mathrm{eff}}\mathbf P] +\frac{1}{2}\mathrm{Tr}\!\left[ \left(\mathbf J^{\mathrm{eff}} -\frac{1}{2}\mathbf K^{\mathrm{eff}}\right)\mathbf P \right]\\ &+2\xi A+\omega_c\xi^2+E_{\mathrm{nuc}}, \end{aligned} \label{eq:static_CS_main} \end{equation} where $A=\mathrm{Tr}[\mathbf g\mathbf P]$, $\xi=-A/\omega_c$, and $E_{\mathrm{nuc}}$ is the nuclear repulsion energy. The coherent photon and light-matter contributions are therefore counted once through $E_{\mathrm{stat}}^{\mathrm{CS}}$.} \red{To write the remaining contributions~\cite{Galitskii1958,saekkinen2015manybody}, we define the one-spin contractions \begin{equation} C_{\alpha} = -\int_0^\beta d\tau\, \mathrm{Tr}\!\left[ \mathbf{\Sigma}^{\alpha}(\tau) \tilde{\mathbf G}(\beta-\tau) \right], \qquad \alpha=ee,eb. \label{eq:Cee_Ceb_main} \end{equation} The electronic correlation contribution~\cite{Galitskii1958} is $C_{ee}$. The residual electron-boson interaction energy~\cite{saekkinen2015manybody} is
\begin{equation}
\begin{aligned}
    E_{eb}^{\mathrm{fluc}} = 2C_{eb}
&=
-2\int_0^\beta d\tau\,
\Tr_e\!\left[
\tilde{\mathbf\Sigma}^{eb}(\tau)
\tilde{\mathbf G}(\beta-\tau)
\right] \\
&= -\int_0^\beta d\tau\,\Pi(\tau)D(\tau).
\end{aligned}
\label{eq:Eeb_main} 
\end{equation} The two forms in Eq.~\eqref{eq:Eeb_main} follow from the same Fan self-energy and polarization and provide an internal consistency relation.} \red{The bosonic fluctuation energy is obtained from the equal-time displacement propagator and the stationary bosonic virial identity~\cite{theophilou2020virial}, \begin{equation} E_b^{\mathrm{fluc}} = -\frac{\omega_c}{2}\left[D(0^-)+1\right] +\frac{1}{2}E_{eb}^{\mathrm{fluc}}. \label{eq:Eb_main} \end{equation} The photon zero-point energy is omitted, consistently with Eq.~\eqref{eq:H_QED_final}. Combining these contributions gives \begin{equation} \begin{aligned} E_{\mathrm{QED-GF2}} &= E_{\mathrm{stat}}+C_{ee} +E_{eb}^{\mathrm{fluc}}+E_b^{\mathrm{fluc}}\\ &= E_{\mathrm{stat}}+C_{ee}+3C_{eb} -\frac{\omega_c}{2}\left[D(0^-)+1\right]. \end{aligned} \label{eqedgf2_total} \end{equation} For CS-GF2, $E_{\mathrm{stat}}$ is given by Eq.~\eqref{eq:static_CS_main}. The same expression applies to the projected LF Hamiltonian when the static energy, interaction tensors, and residual vertex are replaced by their LF counterparts. A detailed derivation is provided in the Supporting Information.} \red{At zero electron-boson coupling, the CS expression reduces to the electronic GF2 energy, with the nuclear repulsion energy added, plus the free thermal photon energy $\omega_c n_B(\omega_c)$. The latter vanishes as $\beta\rightarrow\infty$.}

\red{The coupled total-energy expression and the corresponding self-consistent implementation are independently tested using an exactly solvable two-site electron-boson model, as described in the Supporting Information.}

\paragraph{QED-GF2 Self-Consistent procedure}

The self-consistent procedure used in this work is summarized as follows
\begin{enumerate}
    \item Start from a converged QED mean-field solution, defining $\tilde{\mathbf{F}}$, $\tilde{E}$, and an initial $\tilde{\mathbf{G}}$ with $\tilde{\mathbf{\Sigma}}=0$.
    \item Given $\tilde{\mathbf{G}}$, construct the electronic GF2 self-energy $\mathbf{\Sigma}^{ee}$ from Eq.~\eqref{eq:Sigma_GF2}.
    \item Given $\tilde{\mathbf{G}}$, construct the bosonic polarization $\Pi$ from Eqs.~\eqref{eq:Pi_time_qed} and~\eqref{eq:Pi_freq_qed} and update $D$ from the bosonic Dyson equation Eq.~\eqref{Bosonic Dyson}.
    \item Given $\tilde{\mathbf{G}}$ and $D$, construct the electron-boson self-energy $\mathbf{\Sigma}^{eb}$ from Eqs.~\eqref{eq:Sigma_eb_time_qed} and~\eqref{eq:Sigma_eb_freq_qed}.
    \item Form $\tilde{\mathbf{\Sigma}}=\mathbf{\Sigma}^{ee}+\mathbf{\Sigma}^{eb}$ from Eq.~\eqref{total_sigma} and update $\tilde{\mathbf{G}}$ from the electronic Dyson equation Eq.~\eqref{eq:Dyson_total}.
    \item \red{Update the density matrix from Eq.~\eqref{eq:P_from_G}, rebuild the QED-dressed Fock matrix, update the stationary coherent displacement, and adjust the chemical potential to maintain the required electron number.}
    \item \red{Evaluate the total energy from Eq.~\eqref{eqedgf2_total} and repeat the coupled updates until self-consistency is reached. Both electronic and bosonic Dyson residuals are required to distinguish convergence of the coupled equations from convergence of the energy alone.}
\end{enumerate}
\red{For LF-GF2, this coupled procedure is performed at fixed $\{\lambda_i\}$. The converged correlated energy is then used in the outer optimization of the LF parameters, as described below.}

\subsection{Hamiltonian Transformations and Reference States}
\red{
The coherent-state and Lang-Firsov transformations are unitary transformations and therefore do not depend on the choice of electronic or bosonic reference. We therefore introduce these transformations and their corresponding transformed Hamiltonians before specifying the reference used to evaluate them.
}

\subsubsection{CS-GF2}
Coherent-state transformation applies a real coherent origin shift, $\xi$, to the harmonic oscillators by applying the following unitary transformation
\begin{equation}
    \label{UCS}
    \hat{U}_\text{CS} = e^{-\xi (b - b^\dagger)}, \qquad \xi \in \mathbb{R},
\end{equation}
which displaces the bosons
\begin{equation}
    \hat{U}_\text{CS}^{\dagger} \: b \: \hat{U}_\text{CS} = b + \xi.
\end{equation}
Thus, after applying the unitary transformation in Eq.~\eqref{UCS}, one can write the \emph{coherent-state Hamiltonian} as
\begin{equation}
    \label{UCS On Hamiltonian}
    \hat{H}_{\text{CS}} = \hat{H} + \omega_c \xi^2 + \omega_c \xi (b + b^\dagger) + \sum_{ij} 2 \xi g_{ij}a_i^\dagger a_j.
\end{equation}

\subsubsection{LF-GF2}
Instead of just applying the CS transformation which results in one variational parameter, one can perform the \emph{Lang-Firsov} (LF) transformation~\cite{lang1963kinetic} with variational parameters $\lambda_i$~\cite{Cui2024} defined as
\begin{equation}
    \label{LF Transformation}
    \hat{U}_{\mathrm{LF}} = \exp \left( {\sum_{i}\lambda_i a^\dagger_i a_i \left( b - b^\dagger \right)} \right).
\end{equation}
So we can write an LF-transformed Hamiltonian, $\hat{H}_{\text{LF}}$, by applying both CS and LF transformations~\cite{Cui2024} such that the overall wavefunction can be variationally optimized using parameters $\xi$ and $\lambda_p$
\begin{equation}
    \hat{H}_{\text{LF}} = \hat{U}_{\mathrm{CS}}^\dagger \: \hat{U}_{\mathrm{LF}}^\dagger \: \hat{H} \:\hat{U}_{\mathrm{LF}} \: \hat{U}_{\mathrm{CS}}.
\end{equation}
After applying both transformations~\cite{Cui2024}, the Hamiltonian can be written as
\begin{equation}
\begin{aligned}
\hat{H}_{\text{LF}} =
&\sum_{ij} h_{ij} \, e^{(\lambda_j - \lambda_i)(b - b^\dagger)} a_i^\dagger a_j \\
&+ \frac{1}{2} \sum_{ijkl} v_{ijkl} \, e^{(\lambda_j - \lambda_i + \lambda_l - \lambda_k)(b - b^\dagger)} a_i^\dagger a_k^\dagger a_l a_j \\
&+ \omega_c \left( b^\dagger + \xi - \sum_i \lambda_i a_i^\dagger a_i \right)
\left( b + \xi - \sum_j \lambda_j a_j^\dagger a_j \right) \\
&+ \sum_{ij} g_{ij} \, e^{(\lambda_j - \lambda_i)(b - b^\dagger)} a_i^\dagger a_j \\
&\quad \times \left(
    b + b^\dagger + 2\xi
    - 2 \sum_k \lambda_k a_k^\dagger a_k
\right).
\end{aligned}
\end{equation}

\subsubsection{Reference State}
\red{
Although the preceding QED-GF2 equations are formulated in the finite-temperature Matsubara representation, the applications considered in this work target ground-state properties. 
}
\red{
We therefore introduce the electronic state multiplied by the bosonic vacuum as a ground-state reference in the formal $\beta\rightarrow\infty$ limit. This reference is used to define the mean-field starting point and the parameters entering the transformed Hamiltonian. In the numerical calculations, the subsequent QED-GF2 equations are evaluated at the finite value $\beta=200~E_h^{-1}$, which provides a numerical approximation to the intended ground-state limit.}
\red{Therefore, the chosen wavefunction is}
\begin{equation}
    |\Psi ^{\text{e-b}} \rangle = |\Psi^{\text{e}} \rangle \otimes |0\rangle,
\end{equation}
where $|\Psi^{\text{e}} \rangle$ is the electronic wavefunction and $|0\rangle$ is the boson vacuum state. To incorporate the electron-boson coupling, however, one needs to modify this wavefunction. \red{As introduced in the previous subsection,} \red{we incorporate the coupling effects via two unitary transformations}, namely \emph{Coherent-State} (CS) and \emph{Lang-Firsov} (LF) transformations~\cite{Cui2024,lang1963kinetic}.

\green{Using this ground-state reference, we first evaluate the coherent-state transformed Hamiltonian and construct the corresponding CS-HF reference for the QED-GF2 procedure.}

Therefore, the interacting electron-boson ansatz becomes
\begin{equation}
    \label{CS Ansatz}
    | \Psi ^{\text{CS}} \rangle = \hat{U}_\text{CS} |\Psi ^{\text{e-b}} \rangle.
\end{equation}
Equivalently, this ansatz can be viewed as a unitary transformation of the Hamiltonian~\cite{Cui2024}, Eq.~\eqref{LM Hamil},
\begin{equation}
    \begin{aligned}
        E &= \langle \Psi ^{\text{CS}} | \hat{H} | \Psi ^{\text{CS}} \rangle \\
          &= \langle \Psi ^{\text{e-b}}|
          \underbrace{\hat{U}_{\text{CS}}^\dagger \hat{H} \hat{U}_{\text{CS}}}_{\hat{H}_{\text{CS}}}
          | \Psi ^{\text{e-b}} \rangle \\
          &= \langle \Psi ^{\text{e-b}} | \hat{H}_{\text{CS}} | \Psi ^{\text{e-b}} \rangle.
    \end{aligned}
\end{equation}
Asserting the fact that $\dfrac{\partial E}{\partial \xi} = 0$, the coherent shift $\xi$ can be analytically found~\cite{Cui2024}
\begin{equation}
    \xi = - \sum_{ij} \dfrac{g_{ij} P_{ij}}{\omega_c},
    \label{CS Shift}
\end{equation}
where
\red{$P_{ij}=\sum_\sigma
\langle a_{j\sigma}^\dagger a_{i\sigma}\rangle$}
is the one-electron density matrix.

The modified Fock matrix, CS-HF, can be constructed as~\cite{Cui2024}
\begin{equation}
    \label{CS HF}
    \tilde{F}^{\text{CS}}_{ij}(P, \xi) = F_{ij} (P) + 2 \xi g_{ij}
\end{equation}
and the CS-HF energy is computed using~\cite{Cui2024}
\begin{equation}
    \label{CSHF energy}
    \tilde{E}^{\text{CS-HF}} (P, \xi) = E^{\text{HF}}(P; \xi) + \omega_c \xi^2,
\end{equation}
where $E^{\text{HF}}(P; \xi)$ indicates that the energy is parametrically dependent on $ \xi$.
The modified Fock matrix in Eq.~\eqref{CS HF} is used as the reference for the QED-GF2 procedure, and the coherent shift $\xi$ is updated at each iteration using the current density matrix, defining the resulting method as \emph{CS-GF2}.

\green{Using the same ground-state reference, we then evaluate the LF-transformed Hamiltonian and construct the corresponding effective electronic Hamiltonian.}

To obtain a tractable electronic reference, one can evaluate the expectation value of the LF-transformed Hamiltonian with respect to the boson vacuum state $|0\rangle$. By doing so, we can get an effective electronic Hamiltonian with renormalized one- and two-electron integrals. The complete procedure can be found in Ref.~\cite{Cui2024} to get
\begin{equation}
\label{LF-effective}
\begin{aligned}
\hat{H}^{\text{LF,elec}} 
&= \langle 0 | \hat{H}^{\text{LF}} | 0 \rangle \\
&= \omega_c \, \xi^2 
+ \sum_{ij} \tilde{h}_{ij} \, 
e^{-\frac{1}{2}(\lambda_j - \lambda_i)^2} \, a_i^\dagger a_j \\
&+ \sum_i \left( \omega_c \lambda_i^2 - 2 \omega_c \xi \lambda_i \right)
a_i^\dagger a_i \\
&+ \frac{1}{2} \red{\sum_{ij}} 2 \omega_c \lambda_i \lambda_j \,
a_i^\dagger a_j^\dagger a_j a_i \\
&- \frac{1}{2} \red{\sum_{ijk}} 4 g_{ij} \lambda_k \,
e^{-\frac{1}{2}(\lambda_j - \lambda_i)^2}
\, a_i^\dagger a_k^\dagger a_k a_j \\
&+ \frac{1}{2} \sum_{ijkl} v_{ijkl} \,
e^{-\frac{1}{2}(\lambda_j - \lambda_i + \lambda_l - \lambda_k)^2}
\, a_i^\dagger a_k^\dagger a_l a_j,
\end{aligned}
\end{equation} 
where
\begin{equation}
\tilde{h}_{ij}
= h_{ij} + 2\xi g_{ij} - \lambda_i g_{ij} - g_{ij} \lambda_j.
\end{equation}

The effective Hamiltonian in Eq.~\eqref{LF-effective} can be written in the standard electronic form with renormalized one- and two-electron integrals. To make this connection explicit, we define the effective one- and two-body terms as
\begin{equation}
\begin{aligned}
\tilde{h}^{\mathrm{LF}}_{ij} &= \tilde{h}_{ij} \, 
e^{-\frac{1}{2}(\lambda_j - \lambda_i)^2}
+ \delta_{ij} \left( \omega_c \lambda_i^2 - 2 \omega_c \xi \lambda_i \right).
\end{aligned}
\end{equation}
\begin{equation}
\begin{aligned}
\tilde{v}^{\mathrm{LF}}_{ijkl} &=
v_{ijkl}\,
e^{-\frac{1}{2}(\lambda_j-\lambda_i+\lambda_l-\lambda_k)^2}
\\
&\quad \red{+2\omega_c\lambda_i\lambda_k
\delta_{ij}\delta_{kl}}\\
&\quad \red{-2g_{ij}\lambda_k
e^{-\frac{1}{2}(\lambda_j-\lambda_i)^2}\delta_{kl}}\\
&\quad \red{-2g_{kl}\lambda_i
e^{-\frac{1}{2}(\lambda_l-\lambda_k)^2}\delta_{ij}}.
\end{aligned}
\end{equation}

Using the effective one- and two-electron integrals, one can construct the LF Fock matrix as usual
\begin{equation}
\tilde{F}^{\mathrm{LF}}_{ij} =
\tilde{h}^{\mathrm{LF}}_{ij}
+ \frac{1}{2} P_{kl}
\left( 2 \tilde{v}^{\mathrm{LF}}_{ijkl}
- \tilde{v}^{\mathrm{LF}}_{ilkj} \right)
\end{equation}
where $P_{kl}$ is the one-particle density matrix.

After variational optimization of $\{\lambda_i\}$ and $\xi$, the LF-HF energy can be evaluated via
\begin{equation}
    \label{LF-HF}
    \tilde{E}^{\text{LF-HF}} (P, \xi) = E^{\text{HF}}(P; \xi; \{\lambda_i\}) + \omega_c \xi^2 .
\end{equation}
\red{In the present work, LF-GF2 uses a projected LF Hamiltonian.} \red{The residual linear vertex is \begin{equation} g_{ij}^{\mathrm{LF}} = e^{-\frac{1}{2}(\lambda_j-\lambda_i)^2}g_{ij} -\omega_c\lambda_i\delta_{ij}. \label{eq:gLF_main} \end{equation} Both the Fan self-energy and the bosonic polarization are constructed using $g^{\mathrm{LF}}$, while the electronic second-Born self-energy is constructed using $\tilde v^{\mathrm{LF}}$. Thus, the LF transformation modifies the interaction vertices as well as the Fock reference.} \red{For fixed $\{\lambda_i\}$, the coherent displacement satisfies \begin{equation} \xi = -\frac{\mathrm{Tr}[\mathbf g^{\mathrm{LF}}\mathbf P]} {\omega_c}. \label{eq:xiLF_main} \end{equation} The static LF contribution to Eq.~\eqref{eqedgf2_total} is \begin{equation} E_{\mathrm{stat}}^{\mathrm{LF}} = \frac{1}{2}\mathrm{Tr}\!\left[ \left(\tilde{\mathbf h}^{\mathrm{LF}} +\tilde{\mathbf F}^{\mathrm{LF}}\right)\mathbf P \right] +\omega_c\xi^2+E_{\mathrm{nuc}}, \label{eq:static_LF_main} \end{equation} evaluated with the correlated density matrix. At $\{\lambda_i\}=0$, the projected integrals and residual vertex reduce to their CS counterparts.}

The LF-transformed Fock matrix can serve as a reference
for the QED-GF2 procedure.
\red{For each trial set of LF parameters $\{\lambda_i\}$,
the coupled electronic and bosonic Dyson equations are
first converged. The coherent shift $\xi$ is updated
using the correlated density matrix during this inner
procedure. The resulting converged total energy is
then used to update $\{\lambda_i\}$ in an outer
optimization.}
This defines the resulting method, namely \emph{LF-GF2}.

\red{The perturbative counterpart of CS-GF2 is obtained by evaluating the second-order energy with the CS-HF electronic reference and the free photon propagator.} Subsequent iterations go beyond the second-order perturbation and incorporate higher-order correlation effects.


\red{For LF-GF2, the quantity denoted as LF-MP2 in this work is the projected second-order energy constructed from the same LF electronic interaction and residual linear vertex.}

The LF-GF2 self-consistent procedure is schematically demonstrated in Fig.~\ref{Schematic Procedure and H2 Figure}(c),
\red{where the LF parameters are optimized using the energy of the converged coupled GF2 solution.}


\section{Results}
\begin{figure}[h!]
\centering
\includegraphics[width=0.91\linewidth]{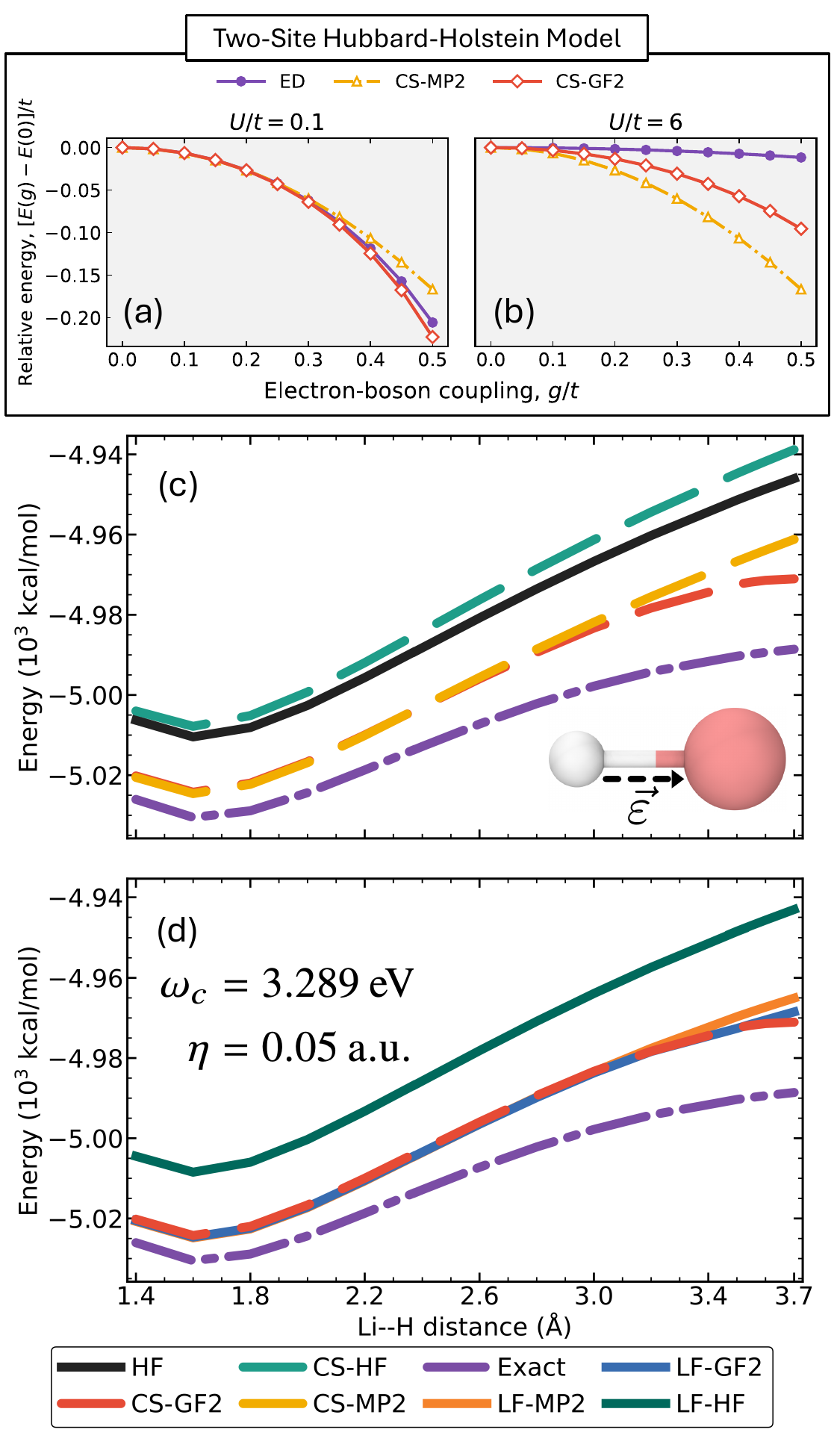}
\caption{\footnotesize
\red{
Comparison of CS-GF2 with reference calculations for the two-site Hubbard-Holstein model and the $\mathrm{LiH}$ molecule.
(a,b) Coupling-induced relative energies, $[E(g)-E(0)]/t$, for the two-site, two-electron Hubbard-Holstein model with $t=\omega_c=1$. Results from exact diagonalization (ED), CS-MP2, and CS-GF2 are shown for (a) $U/t=0.1$ and (b) $U/t=6$.
(c,d) Dissociation curve of $\mathrm{LiH}$ inside an optical cavity with polarization direction $\boldsymbol{\varepsilon}$ along the $\mathrm{Li-H}$ bond, computed using the 6-311G basis set with $\omega_c=3.289~\mathrm{eV}$ and $\eta=0.05~\mathrm{au}$.
(c) Coherent-state results comparing HF, CS-HF, CS-MP2, and CS-GF2 with the exact reference.
(d) Comparison of CS-GF2, LF-HF, LF-MP2, and LF-GF2 relative to the exact result. The LF-MP2 label denotes the projected LF-MP2 energy defined in this work.
}}
\label{LiH inside optical cavity}
\end{figure}
In this section, we assess the performance of the proposed CS-GF2 and LF-GF2 methods against existing QED approaches, including CS-HF, CS-MP2, LF-HF, and LF-MP2.
Additional comparisons are made with FCI and QED-FCI for $\mathrm{H_2}$ and with QED-CASCI for ethylene.
We first compare CS-GF2 with exact diagonalization for a two-site Hubbard-Holstein model. We then consider a range of systems and phenomena: dissociation of diatomic molecules ($\mathrm{H_2}$ and $\mathrm{LiH}$), keto--enol tautomerization barriers and their dependence on cavity coupling strength, the effect of cavity polarization direction on the $\mathrm{(H_2)_2}$ system, and the influence of cavity coupling on the ground-state energy of ethylene ($\mathrm{C_2H_4}$) as a function of the dihedral angle.
\subsection{Computational Details}

\red{
To provide a direct comparison with exact diagonalization, we considered a two-site, two-electron Hubbard-Holstein model Hamiltonian
\begin{equation}
\begin{aligned}
    \hat H_{\mathrm{HH}}
=
&-t\sum_{\sigma}
\left(
c_{1\sigma}^{\dagger}c_{2\sigma}
+
c_{2\sigma}^{\dagger}c_{1\sigma}
\right)
+
U\sum_{i=1}^{2}n_{i\uparrow}n_{i\downarrow}
+
\omega_c b^{\dagger}b\\
&+
g(n_1-n_2)(b+b^{\dagger}),
\end{aligned}
\label{eq:HH_model_main}
\end{equation}
where $t$ is the hopping term, $U$ is the on-site electronic repulsion term, and $g$ is the electron-boson coupling term. We used $t = \omega_c = 1$, and varied the coupling values from $g/t = 0.0 $ to $ 0.5$ for $U/t = 0.1$ and $U/t = 6.0$. Exact diagonalization was performed in the complete two-electron space with bosonic occupations $\mathrm{n} = 0, \cdots, 24$. The CS-GF2 calculations used the same Matsubara representation and self-energy, Dyson-equation, and total-energy implementations as the molecular calculations, with $\beta t=200$. Additional implementation details of the model system calculations are provided in the Supporting Information.
}

CS-HF, LF-HF, MP2, CS-MP2, LF-MP2, GF2, CS-GF2, and LF-GF2 methods were implemented in-house using a custom-developed code. Hartree-Fock and FCI energies, electron integrals, and dipole integrals were computed using \textsc{Psi4}~\cite{10.1063/5.0006002,almlof1982principles,van2006starting,lehtola2019assessment}. QED-CCSD, QED-CASCI and QED-FCI energies were calculated using the $e^{\mathcal{T}}$ program~\cite{10.1063/5.0309334}. \textsc{PySCF}~\cite{sun2015libcint,sun2018pyscf,sun2020recent} was used for the development of an initial CS-HF and CS-MP2 code.
\begin{figure*}[!t]
\centering
\includegraphics[width=1.0\linewidth]{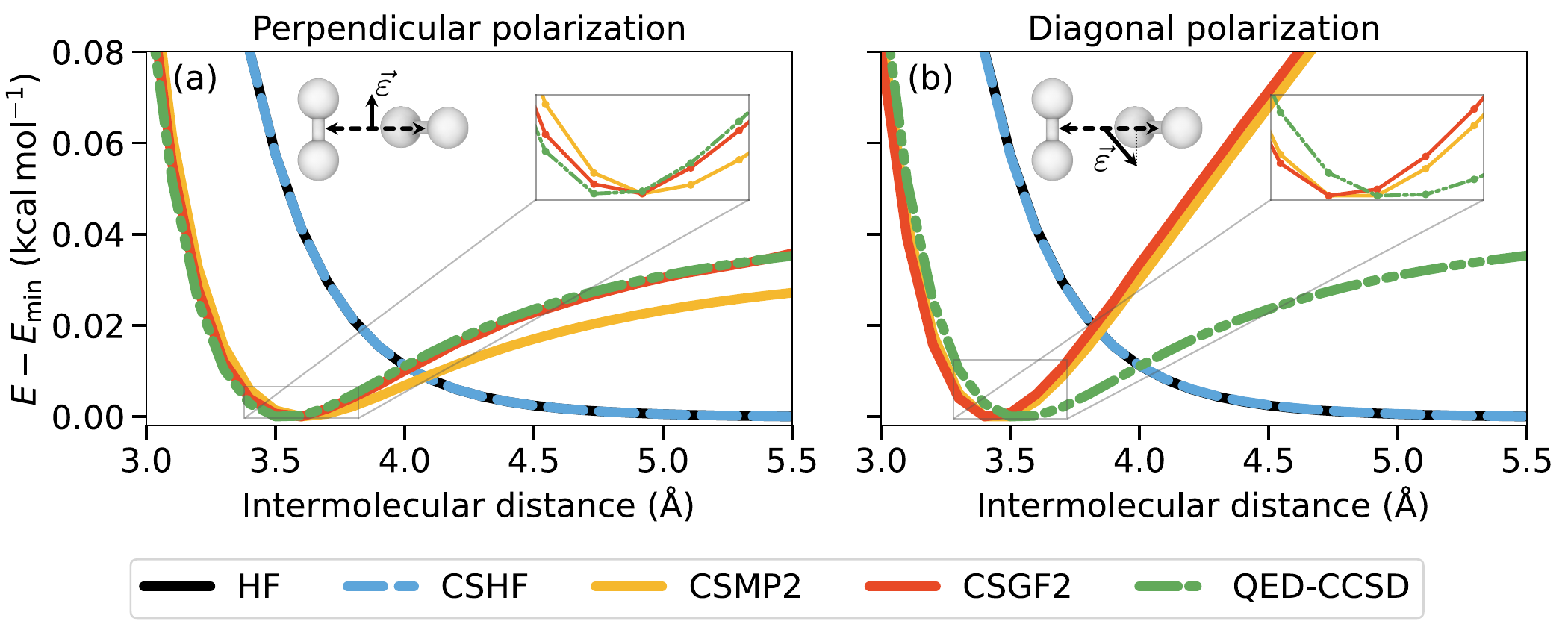}
\caption{\footnotesize
Dissociation curve of a hydrogen dimer in a perpendicular configuration inside an optical cavity with polarization $\varepsilon$ (a) perpendicular to the separation distance and, (b) chosen to have components along all three $x$-, $y$-, and $z$-axes, i.e., $\boldsymbol{\varepsilon} = (1/\sqrt{3},1/\sqrt{3},-1/\sqrt{3})$ computed with the aug-cc-pVDZ basis set for $\omega_c = 2.72~\mathrm{eV}$ and $\eta = 0.01\:\mathrm{au}$.
\red{Energies are reported relative to the minimum of each curve. The insets show the region around the minima.}}
\label{Dimer}
\end{figure*}
We employed a combination of basis sets: cc-pVDZ for $\mathrm{H_2}$, 6-311G for $\mathrm{LiH}$, keto-enol tautomers, and $\mathrm{C_2H_4}$, and aug-cc-pVDZ for the $\mathrm{(H_2)_2}$ system. Optimized geometries for the transition state for keto-enol tautomerization and the equilibrium bond lengths of $\mathrm{C_2H_4}$, were obtained using \textsc{Psi4}~\cite{10.1063/5.0006002}; further details are provided in the Supplementary Information.
For the $\mathrm{(H_2)_2}$ system, the $\mathrm{H_2}$ equilibrium bond length of $0.741  \: \mathrm{\mathring{A}}$ was taken from the Computational Chemistry Comparison and Benchmark Database experimental geometries~\cite{NIST_CCCBDB_2022}.
\red{The present GF2 implementation uses the finite-temperature Matsubara representation as a numerical route to the ground-state limit. All production GF2 calculations were performed at a single finite inverse temperature of $\beta=200~E_h^{-1}$, without extrapolation to $\beta\rightarrow\infty$. The electronic Green's function was represented using 300 explicitly sampled positive fermionic Matsubara frequencies, with a dense low-frequency region followed by a logarithmically spaced high-frequency region. These quantities were interpolated onto a 10,000-point positive-frequency grid for the frequency-to-time transformations. The electronic imaginary-time quantities were represented on a 300-point Chebyshev grid, and the self-energy was transformed using a 50-term Legendre representation.}

\red{We tested convergence with respect to $\beta$ and the Matsubara representation for $\mathrm{H_2}$ at the equilibrium geometry, $R=0.741~\mathrm{\mathring{A}}$, and the stretched geometry, $R=3.000~\mathrm{\mathring{A}}$. For the $\beta$ tests, the full frequency grid and imaginary-time grid were increased approximately proportionally with $\beta$, while the number of explicitly sampled positive frequencies and Legendre terms were kept fixed.}

\red{Across $\beta=50$--$300~E_h^{-1}$, deviations from the production result at $\beta=200~E_h^{-1}$ remained below $0.139~\mathrm{kcal\,mol^{-1}}$ at equilibrium and $0.935~\mathrm{kcal\,mol^{-1}}$ at the stretched geometry, with the latter maximum occurring at the smallest value, $\beta=50~E_h^{-1}$. For $\beta\geq100~E_h^{-1}$, the stretched-geometry deviation was at most $0.106~\mathrm{kcal\,mol^{-1}}$. At fixed $\beta=200~E_h^{-1}$, the coarse and fine Matsubara grids changed the stretched $\mathrm{H_2}$ energy by only $0.039$ and $0.027~\mathrm{kcal\,mol^{-1}}$, respectively. These results establish the numerical stability of the production energies with respect to both $\beta$ and the Matsubara representation. The complete convergence data are reported in Tables~S4 and S5 of the Supporting Information.}

The number of variational parameters $\{ \lambda_i \}$ in the LF methods was chosen to be equal to the number of atomic orbitals. Numerical optimization of these variational parameters was carried out using the limited-memory Broyden--Fletcher--Goldfarb--Shanno (L-BFGS) method.

\subsection{Two-Site Hubbard-Holstein Model}

\red{
Figure~\ref{LiH inside optical cavity}(a,b) compares the relative energies obtained from CS-MP2 and CS-GF2 with exact diagonalization for a two-site Hubbard-Holstein model.
}

\red{
We observe that at $U/t=0.1$, CS-GF2 remains close to the exact result over the plotted coupling range. Moreover, while the difference of ED energy and CS-MP2 continues to increase as the coupling gets higher, the CS-GF2 energies continue to follow the ED energy values closely.
}

\red{
At $U/t=6.0$, we see that CS-MP2 substantially overestimates the energy lowering compared to the ED energy, while the CS-GF2 self-consistency reduces the overestimation within the plotted range. Thus, even though CS-GF2 does not provide a quantitatively exact description of the energy, it reduces the overestimation caused by CS-MP2.
}

\subsection{Cavity Modified Dissociation of Diatomic Molecules}

Figure ~\ref{Schematic Procedure and H2 Figure} shows the dissociation curve of $\mathrm{H_2}$ (d) outside and (e) inside an optical cavity with frequency $\omega_c = 2.99~\mathrm{eV}$ and light-matter coupling strength $\eta = 0.05\:\mathrm{au}$.
In Fig.~\ref{Schematic Procedure and H2 Figure}(d), we find that GF2
\red{removes the downward turn of MP2 at stretched bond lengths and produces a flatter dissociation profile}
~\cite{phillips2014communication,takeshita2019stochastic,rusakov2016self,neuhauser2017stochastic,rosen1931normal}.
\red{However, the FCI reference shows that the relative energies at stretched bond lengths remain overestimated. Thus, flattening of the curve does not by itself establish the correct dissociation limit.}
\red{In Fig.~\ref{Schematic Procedure and H2 Figure}(e), CS-GF2 and LF-GF2 are nearly coincident over the plotted range, and both remain above the QED-FCI relative-energy curve at stretched bond lengths.}
Despite this, LF-GF2 does not provide a substantial improvement over CS-GF2.
Consequently, the additional variational parameters introduced by the Lang-Firsov transformation provide only a minor correction
\red{to the plotted dissociation profile.}

Figure~\ref{LiH inside optical cavity}(c,d) shows the
dissociation curve of the $\mathrm{LiH}$ molecule inside an optical cavity with frequency
\red{$\omega_c = 3.289~\mathrm{eV}$}
and light-matter coupling strength $\eta = 0.05$.
In Fig.~\ref{LiH inside optical cavity}(c), the energy profiles of CS-based methods (CS-HF, CS-MP2, and CS-GF2) are compared against the exact QED energies. We observe that CS-GF2
\red{and CS-MP2 give similar energies near the minimum, while CS-GF2 gives lower energies at larger bond distances.}
In Fig.~\ref{LiH inside optical cavity}(d), LF-based methods are compared against CS-GF2 and the exact QED energies.
\red{LF-GF2 is also close to its first-iteration result near the minimum, with a larger difference appearing toward the stretched end of the scan.}
\red{Both LF-GF2 and CS-GF2 remain above the exact reference, with the difference increasing away from the minimum.}
\red{Consistent with Fig.~\ref{Schematic Procedure and H2 Figure}(e), the CS-GF2 and LF-GF2 profiles are similar, and LF-GF2 does not consistently improve the agreement with the reference.}

In both Figs.~\ref{Schematic Procedure and H2 Figure} and \ref{LiH inside optical cavity}, we observe that implementing the GF2 framework
\red{does not remove the discrepancies with the reference curves at stretched bond lengths.}
\red{LF-GF2 also requires additional optimization,}
as we need to optimize additional variational parameters ${ \lambda_1, \cdots, \lambda_{N_{\text{AO}}} }$ where $N_{\text{AO}}$ is the number of atomic orbitals in each iteration. Therefore, for the later results, we omit LF-GF2 calculations
\red{and focus on CS-GF2, given the small differences between the two GF2 approaches in these comparisons.}

\subsection{Cavity Modified Van der Waals Interactions}
Here we study the van der Waals interactions between two $\mathrm{H_2}$ molecules in a hydrogen dimer in a perpendicular configuration inside an optical cavity with frequency $\omega_c = 2.72~\mathrm{eV}$ and light-matter coupling strength $\eta = 0.01$. In Fig.~\ref{Dimer}(a), the polarization direction $\boldsymbol{\varepsilon}$ is chosen to be orthogonal to the displacement direction, $\boldsymbol{\varepsilon} = (0,0,1/\sqrt{3})$, i.e., perpendicular to the line connecting the center of mass of each $\mathrm{H_2}$ molecule to that of the other. In Fig.~\ref{Dimer}(b), the polarization direction is chosen to have components along all three $x$-, $y$-, and $z$-axes, i.e., $\boldsymbol{\varepsilon} = (1/\sqrt{3},1/\sqrt{3},-1/\sqrt{3})$.

In Fig.~\ref{Dimer}(a)(b), we observe that both HF and CS-HF fail to capture van der Waals interactions, and thus their energy profiles decay monotonically and flatten at large intermolecular separations. This highlights the need for post-Hartree-Fock methods to capture dispersion effects. In Fig.~\ref{Dimer}(a), CS-GF2
\red{closely follows the QED-CCSD relative-energy profile, including the region around the minimum.}
At larger separations,
\red{CS-GF2 remains close to QED-CCSD, while CS-MP2 approaches a lower relative-energy plateau.}

In contrast, in Fig~\ref{Dimer}(b), although CS-MP2 and CS-GF2
\red{show minima in the same general region as QED-CCSD, their relative energies increase much more rapidly as the intermolecular separation increases.}
\red{The comparison therefore reveals a strong polarization dependence of the discrepancy between the CS-based methods and QED-CCSD.}

\begin{figure*}[!t]
\centering
\includegraphics[width=1.0\linewidth]{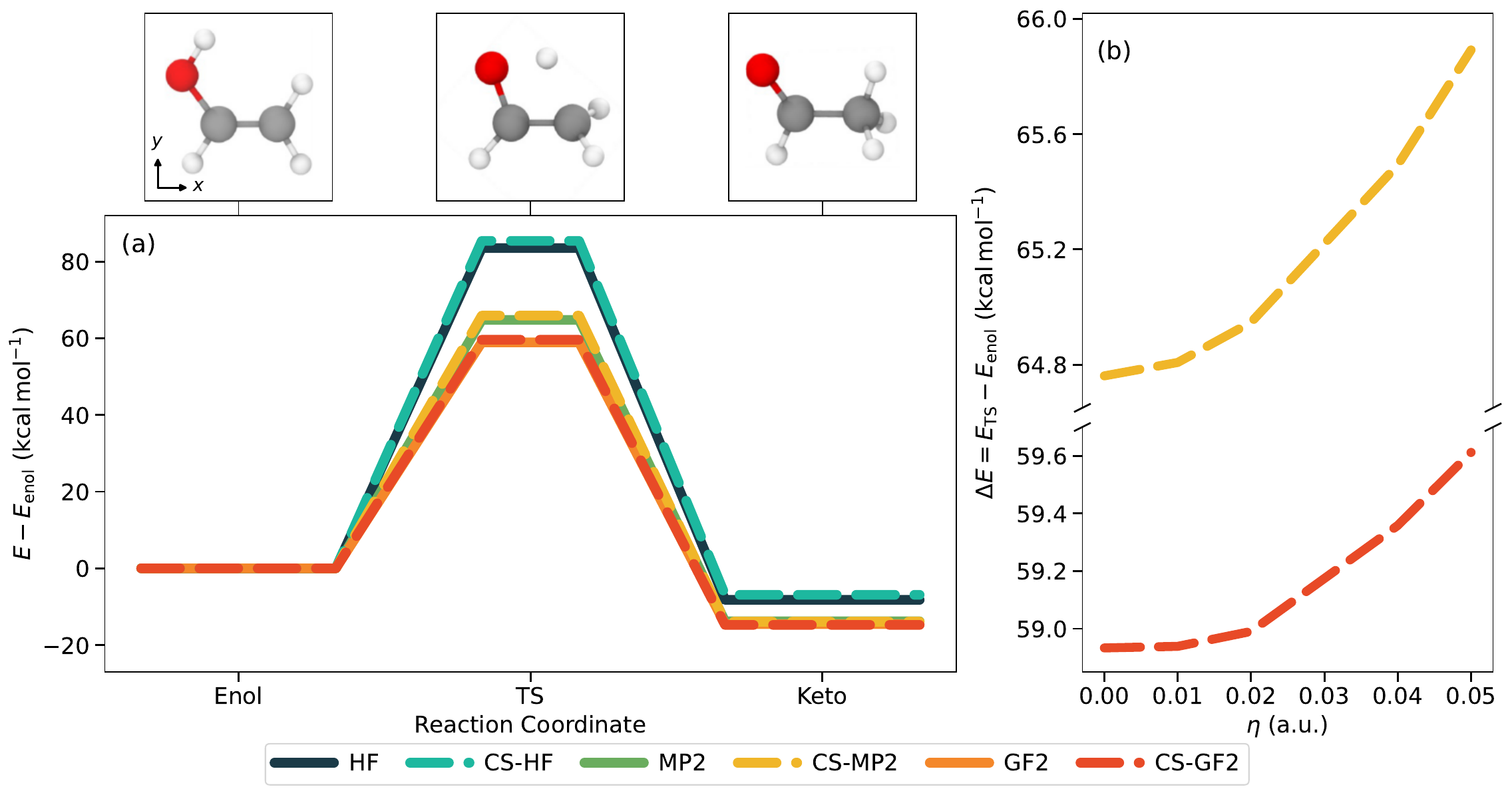}
\caption{\footnotesize (a) Keto–enol tautomerization
\red{relative energy profiles}
(vinyl alcohol $\rightleftharpoons$ acetaldehyde) inside an optical cavity with frequency $\omega_c = 2.72~\mathrm{eV}$, light–matter coupling strength $\eta = 0.05~\mathrm{au}$, and cavity polarization oriented along the $x$-axis, $\boldsymbol{\varepsilon} = (1,0,0)$, computed with the 6-311G basis set.
\red{Energies are reported relative to the enol configuration for each method. Solid lines show the electronic results at $\eta=0$, and dashed lines show the corresponding cavity results.}
(b)
\red{Reaction barrier}
$\Delta E = E_{\mathrm{TS}} - E_{\mathrm{enol}}$ as a function of the coupling strength $\eta$.
\red{The vertical axis is broken to resolve the CS-MP2 and CS-GF2 trends. Lines connect the displayed calculations at $\eta=0.00$, $0.01$, $0.02$, $0.04$, and $0.05~\mathrm{au}$.}}
\label{keto-enol}
\end{figure*}

More fundamentally,
\red{for noninteracting subsystems, an additive reference satisfies}
\begin{equation}
\label{phys meaningful hamiltonian}
H_0 (A + B) = H_0(A) + H_0(B)
\end{equation}
for two subsystems $A$ and $B$.
\red{The size-intensivity problem associated with the CS reference~\cite{ElMoutaoukal2025} is therefore a possible source of the observed discrepancy. However, the present comparison does not separately establish the contributions of the reference and the self-consistent correlation approximation.}

\red{In the plotted range, QED-CCSD approaches a plateau for both polarization directions, whereas CS-MP2 and CS-GF2 show the pronounced upward trend only for the diagonal polarization.}
\red{This behavior identifies a limitation of the present CS-based calculations, while its origin requires a separate assessment of the reference and correlation treatment.}

\subsection{Cavity Modified Keto–Enol Tautomerization}
In Fig.~\ref{keto-enol} we investigate the keto-enol tautomerization
\[
\underbrace{\mathrm{CH_2{=}CHOH}}_{\text{vinyl alcohol}}
\;\rightleftharpoons\;
\underbrace{\mathrm{CH_3CHO}}_{\text{acetaldehyde}}
\]
inside an optical cavity with frequency $\omega_c = 2.72~\mathrm{eV}$ and light-matter coupling strength $\eta = 0.05~\mathrm{au}$. The cavity polarization is oriented along the $x$-axis,
\red{$\boldsymbol{\varepsilon} = (1,0,0)$.}
In Fig.~\ref{keto-enol}(a), the transition state (TS) and keto configuration energies are reported relative to the enol configuration for different methods, and Fig.~\ref{keto-enol}(b) shows the
reaction barrier energy vs. cavity coupling strength.

In Fig.~\ref{keto-enol}(a), both CS-MP2 and CS-GF2 yield higher enol-TS barriers compared to their non-polaritonic counterparts MP2 and GF2. Overall, both (CS)-MP2 and (CS)-GF2 substantially lower the
\red{TS energy relative to the enol minimum compared with the corresponding mean-field results. This lowering reflects the effect of electronic correlation, while the cavity-induced change is obtained by comparing each electronic method with its corresponding CS result.}
Both CS-MP2 and CS-GF2 also predict substantially lower barriers than CS-HF, with CS-GF2 giving a
lower value, approximately
\red{$6.28 \: \mathrm{kcal.mol^{-1}}$}
below
\red{CS-MP2. Lower barriers alone do not establish greater accuracy in the absence of an independent benchmark for this reaction.}
All methods predict the keto configuration to be lower in energy than the enol configuration, with CS-GF2 yielding the largest stabilization,
\red{$\Delta E _{\text{Keto-Enol}} \approx -14.75~\mathrm{kcal.mol^{-1}}$.}

In Fig.~\ref{keto-enol}(b), we examine the dependence of the reaction barrier $\Delta E = E_{\mathrm{TS}} - E_{\mathrm{enol}}$ on the coupling strength $\eta$. At $\eta = 0$, CS-MP2 and CS-GF2 reduce to MP2 and GF2, giving barriers of approximately
\red{$64.76~\mathrm{kcal.mol^{-1}}$ and $58.93~\mathrm{kcal.mol^{-1}}$,}
respectively.
\red{At $\eta=0.05~\mathrm{au}$, the corresponding barriers are approximately $65.89~\mathrm{kcal\,mol^{-1}}$ and $59.61~\mathrm{kcal\,mol^{-1}}$, giving cavity-induced increases of $1.13~\mathrm{kcal\,mol^{-1}}$ and $0.68~\mathrm{kcal\,mol^{-1}}$.}
As the coupling strength increases, both methods predict a monotonic increase in the reaction barrier
\red{over the displayed coupling values},
with CS-GF2 consistently yielding lower barriers than CS-MP2,
\red{indicating that the cavity raises the TS energy relative to the enol minimum.}

\begin{figure*}
\centering
\includegraphics[width=1.0\linewidth]{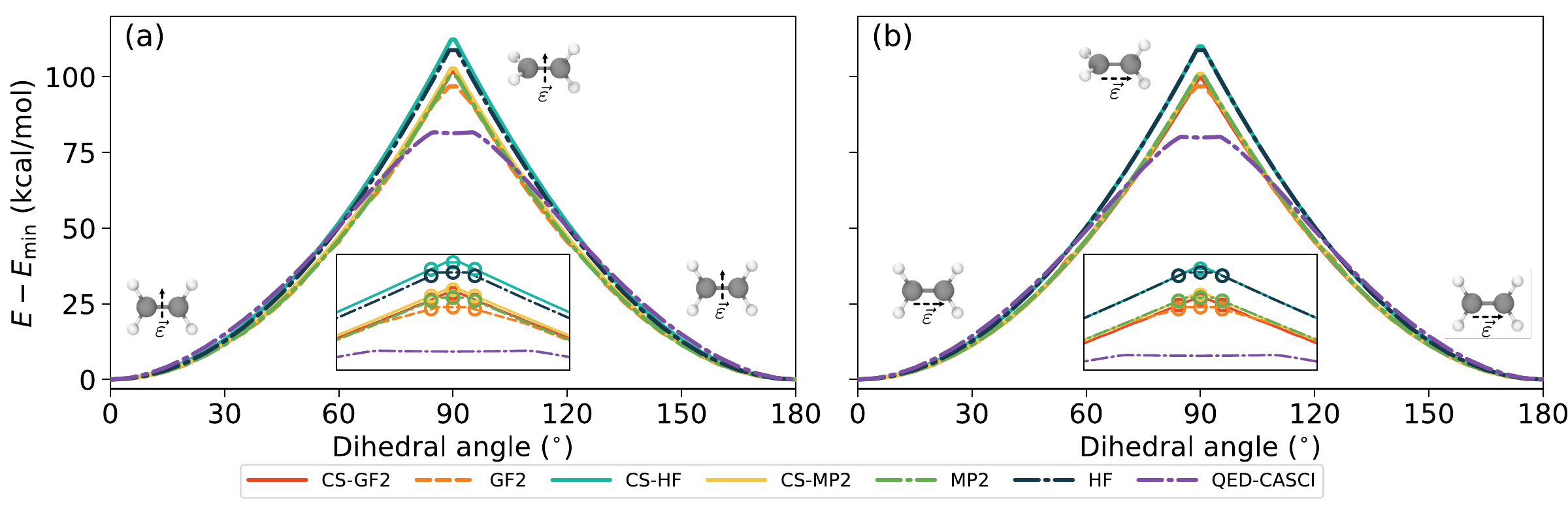}
\caption{\footnotesize Torsional potential energy surface of ethylene molecule ($\mathrm{C_2H_4}$) inside an optical cavity with frequency $\omega_c = 2.72~\mathrm{eV}$, light–matter coupling strength $\eta = 0.05~\mathrm{au}$, and cavity polarization (a) perpendicular to the $\mathrm{C{-}C}$ bond and (b) parallel to the $\mathrm{C{-}C}$ bond, computed with the 6-311G basis set.
\red{The corresponding electronic HF, MP2, and GF2 profiles are included for comparison, together with the QED-CASCI active-space reference. Energies are reported relative to the minimum of each curve. The insets show the region around $\theta=90^\circ$.}}
\label{ethylene}
\end{figure*}
\subsection{Cavity Modified Torsional Potential of Ethylene}
In Fig.~\ref{ethylene} we study the torsional potential energy surface of the ethylene molecule, $\mathrm{C_2H_4}$, inside an optical cavity with frequency $\omega_c = 2.72~\mathrm{eV}$ and light-matter coupling strength $\eta = 0.05~\mathrm{au}$.  The relative energy $E - E_{\text{min}}$ is reported as a function of the dihedral angle $\theta$ between the two $\mathrm{CH_2}$ fragments. In Fig.~7(a), the cavity polarization is oriented perpendicular to the $\mathrm{C-C}$ bond, while in Fig.~7(b) the polarization is parallel to the $\mathrm{C-C}$ bond.
\red{The QED-CASCI curves provide an active-space reference for assessing the torsional profiles, particularly near the strongly correlated twisted configuration.}

In both Fig.~\ref{ethylene}(a) and Fig.~\ref{ethylene}(b), all methods
predict the expected increase in energy as the molecule twists from the planar configuration ($\theta = 0^\circ$) toward the perpendicular geometry ($\theta = 90 ^ \circ$), which corresponds to the breaking process of the $\pi$ bond. Therefore, the planar configurations ($\theta = 0^\circ, 180^\circ$) represent the lowest-energy structures, while the twisted configuration near $\theta = 90 ^ \circ$ corresponds to the torsional barrier of ethylene.
\red{However, the QED-CASCI reference gives a substantially lower and broader barrier region than the GF2 profiles. Thus, the overall increase in energy upon torsion does not by itself establish a quantitatively accurate description near $90^\circ$.}

In Fig.~\ref{ethylene}(a), where the cavity polarization is orthogonal to the $\mathrm{C-C}$ bond, both CS-MP2 and CS-GF2
\red{raise}
the torsional barrier relative to their non-polaritonic counterparts MP2 and GF2. A similar trend is observed at the Hartree-Fock level, where CS-HF yields
\red{higher relative energies near the barrier}
than HF. Among all methods,
\red{QED-CASCI}
predicts the lowest torsional barrier, as highlighted in the inset in Fig.~\ref{ethylene}(a).

In contrast, Fig.~\ref{ethylene}(b) shows that when the cavity polarization is parallel to the $\mathrm{C-C}$ bond, the polaritonic and non-polaritonic energy profiles become much closer to one another. In this case, the cavity induces only a relatively small modification of the torsional barrier, with
\red{each electronic method and its corresponding cavity extension yielding closely spaced energy profiles}
near the twisted configuration around $90^\circ$. This behavior indicates that the cavity-induced modification of the torsional potential strongly depends on the relative orientation between the molecular electronic structure and the cavity polarization direction.
\red{Nevertheless, for both polarization directions, the discrepancy between CS-GF2 and QED-CASCI near $90^\circ$ is larger than the cavity-induced change in the GF2 barrier. The present restricted GF2 treatment therefore does not quantitatively reproduce the active-space reference in this region.}
\subsection{Conclusions}
In this work, we developed a many-body Green's function method for describing the electronic structure of strongly coupled light-matter systems by combining the second-order Green's function approach with coherent state (CS) and Lang-Firsov (LF) transformations. This combination leads to two approaches, namely CS-GF2 and LF-GF2, built on top of the self-consistent GF2 method to include electron-boson-coupled systems, thus providing a practical route to combining the many-body treatment of electronic correlation into \emph{ab initio} QED calculations.

\red{We first performed calculations for a two-site Hubbard-Holstein model. Comparison with exact diagonalization for the two-site Hubbard-Holstein model shows that CS-GF2 closely reproduces the coupling-induced energy change in the weakly interacting regime. In the strongly interacting regime, CS-GF2 reduces the overestimation obtained with CS-MP2, although a remaining discrepancy from the exact result is observed.}
We then benchmarked these methods for the diatomic molecules $\mathrm{H_2}$ and $\mathrm{LiH}$. We observed that both CS-GF2 and LF-GF2
\red{produce changes relative to their first-iteration energies, particularly at stretched bond lengths.}
\red{For $\mathrm{H_2}$, GF2 removes the downward turn of MP2, but the stretched-bond relative energies remain above the FCI and QED-FCI references. For $\mathrm{LiH}$, CS-GF2 and LF-GF2 give similar profiles and remain above the exact reference.}
Although LF-GF2 can slightly improve the results in some cases, the
\red{differences from CS-GF2 are relatively small, while LF-GF2 requires optimization of the Lang-Firsov variational parameters at each iteration.}
These results show that the self-consistent Green's function procedure captures correlation effects beyond MP2 in these systems.
\red{However, the addition of LF variational parameters does not remove the remaining discrepancies with the reference calculations.}

We then applied our newly developed CS-GF2 framework to study cavity-modified intramolecular and intermolecular interactions. Specifically, we examined cavity-modified van der Waals interactions, keto-enol tautomerization, and the torsional potential of ethylene. We observed that cavity-induced changes in the ground-state energy depend on both the molecule's electronic structure and the orientation of the cavity's polarization.

For the hydrogen dimer, our results show that CS-GF2
\red{closely follows the QED-CCSD relative-energy profile}
when the cavity polarization is perpendicular to the separation direction. However, when the polarization has a component along the separation direction, we observe
\red{a pronounced discrepancy from QED-CCSD at larger separations.}
For separated molecular fragments, the long-range behavior can depend sensitively on the polarization direction and on the size-intensivity of the reference ansatz. This observation underscores the importance of choosing an appropriate ansatz to describe specific molecular interactions.
\red{The present comparison does not establish whether the reference or the correlation approximation is primarily responsible for this discrepancy.}
For the keto-enol tautomerization, we
\red{examine}
the applicability of our method to real molecular reactions and observe that the (CS-)GF2 method predicts a lower barrier than (CS-)MP2, a trend observed across a range of light-matter coupling values.
\red{At $\eta=0.05~\mathrm{au}$, the cavity increases the enol-TS barrier by approximately $0.68~\mathrm{kcal\,mol^{-1}}$ within CS-GF2 and $1.13~\mathrm{kcal\,mol^{-1}}$ within CS-MP2.}
The ethylene torsional scan demonstrates the polarization direction dependence on the torsional energy profile,
\red{but the comparison with QED-CASCI reveals a substantial remaining discrepancy near $90^\circ$. This region therefore identifies a limitation of the present restricted GF2 treatment.}

Overall, our work establishes CS-GF2 and LF-GF2 as systematically improvable many-body treatments for ground-state molecular polaritons. Our results suggest that the self-consistent GF2 procedure provides
\red{a framework}
for computing the electronic structure of correlated light-matter systems.
\red{The reference comparisons also identify regimes where further improvement of the correlation treatment is needed.}
However, the computational cost of CS-GF2 remains a bottleneck. In the future, we will focus on lowering the computational cost of our method as well as extending the (QED-)GF2 formalism to open-shell systems.

\section {Data Availability}
The data supporting the results of this study are publicly available via the GitHub repository in Ref.~\cite{Data_GitHub}.
\section {Code Availability}
The source code supporting the findings of this study is available upon reasonable request.
\section {Acknowledgments}
This work was supported by Texas A\&M startup funds and the Strategic Transformative Research Program (STRP) grant provided by Texas A\&M University. This work was partially supported by  the U.S. National Science Foundation (NSF) under Grant
No. CHE-2611431. This work utilized  TAMU LAUNCH at Texas A\&M University, and SDSC Expanse at the San Diego Supercomputer Center through allocation CHE250156 from the Advanced Cyberinfrastructure Coordination Ecosystem: Services\cite{boerner2023ACCESS} \& Support (ACCESS) program, which is supported by U.S. National Science Foundation grants \#2138259, \#2138286, \#2138307, \#2137603, and \#2138296. LM acknowledges support from ANID FONDECYT Iniciación Grant No. 11250602. LM and JC acknowledge the supercomputing infrastructure of the National Laboratory for High Performance Computing (NLHPC, CCSS210001), which supported part of the test calculations reported in this work.



\section{Competing Interests}
The authors declare no competing interests.

\bibliography{bib.bib}
\end{document}